\documentclass[pre,aps,12pt,superscriptaddress]{revtex4-1}

\usepackage{graphicx}
\usepackage{appendix}
\usepackage{amsmath,amssymb}

\newcommand{\degree}{\ensuremath{^\circ}}
\newcommand{\ignorar}[1]{}

\begin{document}

\title{Lagrangian transport in a microtidal coastal area: the Bay of Palma, island of Mallorca, Spain}

\author{Ismael Hern\'andez-Carrasco} 
\author{Crist\'obal L\'opez}
\author{Emilio Hern\'andez-Garc\'ia} 
\affiliation{
IFISC, Instituto de F\'isica Interdisciplinar y Sistemas Complejos (CSIC-UIB),
07122 Palma de Mallorca, Spain}

\author{Alejandro Orfila} 
\affiliation{IMEDEA, Instituto Mediterr\'aneo de Estudios
Avanzados (CSIC-UIB), 07190 Esporles, Spain}

\date{\today}

\begin{abstract}

Coastal transport in the Bay of Palma, a small region in the
island of Mallorca, Spain, is characterized in terms of
Lagrangian descriptors. The data sets used for this study are
the output for two months (one in autumn and one in
summer) of a high resolution numerical model, ROMS, forced
atmospherically and with a spatial resolution of 300 m.
The two months were selected because its different wind
regime, which is the main driver of the
sea dynamics in this area. Finite-size Lyapunov Exponents (FSLEs) 
 were used to locate semi-persistent
Lagrangian coherent structures (LCS) and to understand the
different flow regimes in the Bay. The different wind
directions and regularity in the two months have a clear impact
on the surface Bay dynamics, whereas only topographic features
appear clearly in the bottom structures. The fluid interchange
between the Bay and the open ocean was studied by computing
particle trajectories and Residence Times (RT) maps. The escape rate of
particles out of the Bay is qualitatively different,
with a 32$\%$ more of escape rate of particles to the
ocean in October than in July, owing to the different
geometric characteristics of the flow. We show that LCSs
separate regions with different transport properties by
displaying spatial distributions of residence times on synoptic
Lagrangian maps together with the location of the LCSs.
Correlations between the time-dependent behavior of FSLE and
RT are also investigated, showing a negative
dependence when the stirring characterized by FSLE values moves
particles in the direction of escape.

\end{abstract}

\maketitle

\section{Introduction}
\label{Sec:Intro}

The study of transport and mixing in coastal flows is of major
interest because of their economic and ecological importance.
Due to the particularities that they present, like influence of
complex topography, coastline shape and the direct driving at
the surface by highly variable wind forcing, coastal flow
dynamics remains still poorly understood.

Recently, coastal observations and modeling efforts in
different regions have been addressed from the Lagrangian point
of view: \citet{Lekien2005} showed that Lagrangian Coherent
Structures (LCSs) computed from velocity fields obtained from
HF Radar measurements can be used to predict pollutant
dispersion in the coast of Florida; \citet{Gildor2009} and
\citet{Shadden2009} detected LCSs with HF Radar data in the
Gulf of Eliat, Israel, and in Monterey Bay, respectively.
\citet{Haza2010} studied small-scale properties of dispersion
measurements obtained from HF Radar data in the Gulf of La
Spezia, Italy. Also \citet{Nencioli2011} have detected LSCs in a coastal 
region with a Lyapunov method based on in situ observations.
Besides Radar measurements, LCSs obtained from
velocity data of high resolution numerical models have been
used to analyze the effect of the waves on LCS in the Bay of
Palma, Spain \citep{Galan2012}, to study the transport in the
tidal flow of Ria de Vigo, Spain \citep{Huhn2012}, or to study
the water quality of a very small coastal region, the Hobie
Beach, USA \citep{Fiorentino2012} Also, data from drifters
released in the Santa Barbara Channel were used by
\cite{Ohlmann2012} to characterize relative dispersion, very
useful to improve Lagrangian stochastic models. The application
of Lagrangian techniques to study the dynamics in a shallow
lake (small closed basin) has been performed in
\cite{Pattantyus2008}.

Palma is the largest city in the Balearic Islands.
Human activities, in particular recreational ones, give to
water quality in the Bay of Palma a large economic value. A
proper analysis of transport can be useful to understand the
fluid dynamics in the Bay and therefore help protect the
coastal water. Previous studies performed in the Bay of Palma
used Eulerian techniques to understand the coastal dynamics
\citep{Jordi2009, Jordi2011}. In this work we study some
transport properties in the Bay of Palma using Lagrangian
techniques developed from dynamical systems theory. Computing
both LCSs and residence times the Bay of Palma can be sorted in
regions of different properties, for example having more or
less connectivity with the open ocean. This kind of studies
have demonstrated to be useful to identify pollution pathways
or conditions for red tides \citep{Lekien2005,Fiorentino2012}.
The Bay is a semi-enclosed basin located in the southwest of
the island of Mallorca (western Mediterranean sea),
whose coastal flow is mainly induced by the wind
\citep{Jordi2009,Jordi2011}. Forcing by tides is almost
negligible with a tidal amplitude of less than 0.25$m$. This
makes the dynamics here different form other locations (e.g.
\citet{Shadden2009,Huhn2012}) where tides are dominant, and
then provides the opportunity to test the performance of
dynamical systems tools in this situation in which forcing only
acts directly on the sea surface, and in which there are rather
different forcing regimes depending on the season. The
Lagrangian diagnosis will be obtained from velocity data of a
realistic numerical model at high resolution, which resolves
spatial scales of a few hundred of meters. We investigate the
surface horizontal transport during two months corresponding to
different seasons (autumn and summer), and therefore to
different wind conditions, in order to highlight the effect 
of the wind on transport. In the case of July we also study the deepest 
bottom layer. We compute the barriers and avenues to transport
(LCS) from lines of high values of Finite-Size Lyapunov
Exponents (FSLE). We also present calculations of residence
times and show synoptic Lagrangian maps (SLM) of these times
\citep{Lipphardt2006}, which will allow us a detailed
visualization of the interchange of fluid particles between the
Bay and the open sea. The relationship between LCSs and areas
of different residence times will be analyzed.

The organization of this paper is as follows. The data set used
in the computations and the area of study is described in
Section \ref{sec:data}. Section \ref{sec:metodo} presents a
brief overview of the Lagrangian tools that are used. Before
presenting the Lagrangian results, we show in Section
\ref{sec:vel} a short summary of Eulerian results by studying
the velocties in the Bay. We present in Section
\ref{sec:results} a characterization of stirring in the Bay of
Palma in terms of FSLE and residence times. Using the
definition of LCS given in Section \ref{sec:metodo}, Lagrangian
barriers are identified in the domain of interest. We compute
escape rates and residence times of fluid particles to describe
the transport relation between the Bay and the open ocean. We
provide possible mechanisms to explain differences in the
residences times and FSLE between different seasonal months.
Finally we summarize the main results in Section
\ref{sec:conclusion}.

\begin{figure*}[htb]
\includegraphics[width=0.90\textwidth]{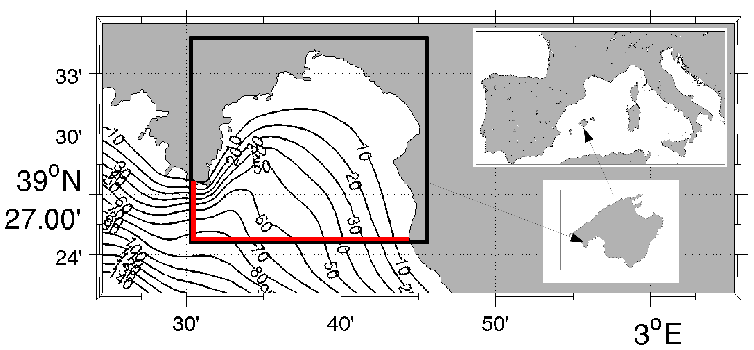}
\caption{Bathymetry contours (in meters) of the model domain.
The black box indicates the Palma Bay and the inset graphics
give the geographical location of Mallorca Island in the
western Mediterranean Sea.
 }
\label{fig:location}
\end{figure*}

\section{Data and characteristics of the study region}
\label{sec:data}
\subsection{Area of study}
\label{sec:area} The island of Mallorca (Fig.
\ref{fig:location}) is part of the Balearic Islands Archipelago
and is located in the center of the western Mediterranean
(between 39$^{\circ}$ and 40$^{\circ}$N and 2.50$^{\circ}$ and
3.50$^{\circ}$ E).
The Bay of Palma is a nearly semi-circular and semi-enclosed
basin located in the southwest coast of Mallorca and it can
reach depths of more than 60 $m$. The Bay of Palma
is defined as the water mass inside the square in
Fig.\ref{fig:location}, consisting of a northern limit at
39$^{\circ}$34$'$N, a southern limit at 39$^{\circ}$24$'$N, and
2$^{\circ}$30$'$E and 2$^{\circ}$45$'$E as the western and eastern
limits, respectively. The open boundary to the sea is
in the southern part and it is 20 km wide.

The size of the Bay is smaller than the Rossby radius of
deformation at these latitudes, and the main circulation is
determined by the bathymetry at the bottom layer and by
local and remote winds at the surface layer. In
particular the studies by \cite{Jordi2009,Jordi2011} have shown
that the major forcing mechanisms come from wind-induced island
trapped waves (ITW) propagating at an island scale and by
locally wind-induced mass balance. The intense ITW can produce
new instabilities which can generate coastal gyres at
submesoscale (see \citet{Jordi2011}). During summer there are
persistent sea breeze conditions. In July and August, the
weather is often almost identical from one day to the next. In
the vicinity of the Bay and along the southern coast of
Mallorca the breeze blows from the south-west. Several studies
\citep{Ramis1988,Ramis1995}, have pointed out that the
meteorological conditions of Mallorca (intense solar radiation,
clear skies, soil water deficit, dryness, weak surface pressure
gradients, etc.) favors the development of sea breeze, often
from April to October, and almost every day during July and
August. Winds in autumn, and particularly in late
September and October are more irregular, with episodes of
strong storm activity \citep{Tuduri1997}.

\subsection{Data}

The velocity data sets were obtained from the numerical model
ROMS (Regional Ocean Model System). ROMS is a free surface,
hydrostatic, primitive equation ocean model. The model
uses a stretched, generalized nonlinear coordinate system to
follow bottom topography in the vertical, and orthogonal
curvilinear coordinates in the horizontal
\citep{Song1994,Haidvogel2000}. At each grid point,
horizontal resolution $\Delta_0$ is the same in both the
longitudinal, $\phi$, and latitudinal, $\theta$, directions.

We run the simulation  with a resolution of $\Delta_0
=0.0027^{\circ}$ ($\sim$300$m$, ROMS300), which is itself
nested into a larger and coarser grid with $\Delta_0$
=1/74$^{\circ}$ ($\sim$1500$m$). Boundary conditions
for the coarser domain were taken from daily outputs of the
Mediterranean Forecasting System \citep{Dobricic2007,Oddo2009}. 
The ROMS300 domain covers $39^{\circ}$12$'$N -
$39^{\circ}$36$'$N (latitude), and $2^{\circ}$24$'$E -
$3^{\circ}$6$'$E (longitude).
The total number of grid nodes is 260 $\times$ 148. Vertical
resolution is variable with $10$ layers in total. All domains
were forced using realistic winds provided by the PSU/NCAR
mesoscale model MM5. The initial vertical structure of
temperature and salinity was obtained from the Levitus database
\citep{Locarnini2006,Antonov2006}.

We will manage velocity data from the surface layer and the
bottom layer for the grid of $\Delta_0$ $\approx$ 300$m$. This
domain allows us to analyze the fluid interchange between the
Bay and the open ocean, using a high resolution velocity field.
Only horizontal velocities are considered, so that
vertical displacements are neglected in the surface layer, and
particles in the bottom remain in the bottom layer. This is
justified by the small integration times we will use.
Nevertheless, close to the coast they can have an impact that
will be the subject of future work. The output of the model was
compared with data from drifters (see \cite{Galan2012}) and a
reasonable agreement was found, although it improved when
adding the influence of wave intensity. Thus the present study
should be considered as a simplified baseline case against
which to compare the future consideration of the full 3d
dynamics, or the influence of small scale process such as waves
\citep{Galan2012}. We will study two different intervals of
time corresponding to two different wind regimes: one
starting on October $5$th, $2008$ and finishing on October
$29$th, $2008$; and the other extending from July $1$st, $2009$
until July $26$th, $2009$. Temporal resolution is 15 minutes
and 10 minutes for October and July, respectively, resulting in
a total of $2375$ snapshots of the velocity field for October,
and $3744$ for July.

\section{Methodology}
\label{sec:metodo}

\subsection{LCSs and particle dispersion from FSLE}
\label{fsle}

Our methodology is based on the Lagrangian analysis of marine
flows. In the Lagrangian view, particles are advected by the
flow and their horizontal motion (neglecting motions between
model layers) is governed by the differential equations
\begin{eqnarray}
\frac{dx}{dt}&=&{v_x(x,y,t)}, \label{eqsmotiona}\\
\frac{dy}{dt}&=&{v_y(x,y,t)},
\label{eqsmotionb}
\end{eqnarray}
where ($x(t),y(t)$) are the west-east and the south-north
coordinates of the trajectories and ($v_x,v_y$) are the
eastwards and northwards components of the velocity. Because of
the small sizes involved, we will use a Cartesian coordinate system.

LCSs \citep{Haller2000b,dOvidio2004,Shadden2005}, are roughly
defined as the material lines organizing the transport in the
flow. They are the analogs, for time-dependent flows, of the
unstable and stable manifolds of hyperbolic fixed points. Among
other approaches \citep{Mancho2006b,Mendoza2010,Mezic2010,Rypina2011,Haller2012},
ridges of the local Lyapunov Exponents provide a convenient
tool to locate them. In our case, we use the so-called
Finite-Size Lyapunov Exponents (FSLEs) which are the adaptation
of the asymptotic classical Lyapunov Exponent to finite spatial
scales \citep{Aurell1997,Boffetta2001}. FSLEs are a local
measure of particle dispersion and thus of stirring and mixing,
as a function of the spatial resolution, serving to isolate the
different regimes corresponding to different length scales of
the oceanic flows, very useful in coastal systems
\citep{libroangelo}. In fact the first applications of
the FSLE technique in oceanography were for closed or
semi-closed basins \citep{Buffoni1996,Buffoni1997}.

For two particles of fluid, one of them located at
$\textbf{x}$, the FSLE at time $t_{0}$ and at the spatial point
$\textbf{x}$ is given by the formula:
\begin{equation}
\lambda (\textbf{x}, t_{0}, \delta_0, \delta_f)= \frac{1}{|\tau|}
\ln{\frac{\delta_f}{\delta_0}},
\label{formFSLE}
\end{equation}
where $\delta_0$ is the initial distance of the two given
particles, and $\delta_f$ is their final distance. Thus, to
compute the FSLEs we need to calculate the minimal time, $\tau$, needed
for the two particles initially separated $\delta_0$, to get a
final distance $\delta_f$ (in this way the FSLE represents the
inverse time scale for mixing up fluid parcels between length
scales $\delta_{0}$ and $\delta_{f}$). To obtain this time we
need to know the trajectories of the particles (from Eqs.
(\ref{eqsmotiona}) and (\ref{eqsmotionb})) which gives the
Lagrangian character to this quantity. 
The FSLEs are computed
for the points $\textbf{x}$ of a square lattice with lattice spacing
coincident with the initial separation of fluid particles
$\delta_{0}$. 
We can obtain a good estimation of the minimal $\tau$ at each site by 
selecting the trajectory which diverges the first among the 
four trajectories starting at the neighbors of the given site in the 
grid of initial conditions. Numerically we integrate the equations of motion
using a standard, fourth-order Runge-Kutta scheme, with an
integration time step corresponding to the time
resolution of the velocity data: $dt=15$ minutes in October
and $dt=10$ minutes in July. 
We have checked in selected trajectories that using in July 
the same time step $dt=15$ as in October does not alter the trajectories.
Since velocity information is
provided just in a discrete space-time grid, spatiotemporal
interpolation of the velocity data is achieved by bilinear
interpolation. For the spatial scales that define FSLEs, we
take $\delta_f=0.1 \degree$, i.e., final separations of about
$10 \ km$, because of the size of the Bay. On the other side,
we take $\delta_0$ equal to $75 \ m$, four times smaller than
the resolution of the velocity field, $\Delta_0 = 300 \ m$.
Since we are interested only in fast time scales, our
integrations are restricted to 5 days. Locations for which the
final separation at the end of this period has not reached the
prescribed $\delta_f=10 \ km$ (or for which particles have been
trapped by land) are assigned a value $\lambda=0$.

FSLEs can be computed from trajectory integration
$\emph{backwards}$ and $\emph{forward}$ in time. Their highest
values as a function of the initial location, \textbf{x},
organize in filamental structures approximating relevant
manifolds: ridges in the spatial distribution of backward
(forward) FSLEs identify regions of locally maximum compression
(separation), approximating attracting (repelling) material
lines or unstable (stable) manifolds of hyperbolic
trajectories, which can be identified with the LCSs
\citep{Haller2000b,dOvidio2004,Shadden2005,TewKai2009,HernandezCarrasco2011},
and characterize the flow from the Lagrangian point of view
\citep{Joseph2002,Koh2002}. Attracting LCSs associated to
backward integration (the unstable manifolds) have a direct
physical interpretation
\citep{Joseph2002,dOvidio2004,dOvidio2009}. Tracers
(chlorophyll, temperature, ...) spread along these attracting
LCSs, thus creating their typical filamental structure
\citep{TelGruiz2006,Lehan2007,TewKai2009,Calil2010}. When not
stated explicitly, by FSLE we will mean the \textsl{backwards}
FSLE values. In addition to locate spatial structures,
time-averages of FSLE give an indication of the intensity of
stirring in given areas, which we analyze in Sect.
\ref{stirring_mallorca}.

We close this section by noting that the relationship between
LCSs and Lyapunov exponents is based on heuristic arguments
which may not be correct in some cases (see for example
\cite{Haller2011a}). We identify as possible LCSs only
the locations having the largest values of FSLE, which align in
linear structures. In this way we effectively select only the
highest FSLE ridges which are more likely to organize the flow.
Even in this case, it is possible that the FSLE technique
identifies regions of high shear which are not hyperbolic and
then may lack some of the properties of \textsl{bona fide}
LCSs. Thus, direct inspection of particle trajectories and
comparison with complementary techniques would be needed to
confirm the validity of the FSLE approach in this situation.
One of such complementary techniques is residence time maps
that we present in the following section.

\subsection{Escape and residence times}
\label{sec:methods_ERSLM}

Another characteristic time-scale for transport processes in
open flows is the so-called escape rate \citep{Lai2011}. This
quantity measures how quickly particles trajectories escape
from a domain. If we initiate $N(0)$ particles in a flow, we
can measure how the trajectories escape the preselected region.
In the case in which the decay in the number of particles
remaining in the region up to time $t$, $N(t)$, decays
exponentially with time, $N(t)/N(0) \sim e^{-\kappa t}$, there
is a well-defined escape time defined as the inverse of the
escape rate $\kappa$: $\tau_e=1/\kappa$. For the range of times
explored in our work, we will see that the particle escape is
close to exponential and then we can estimate the value of
$\tau_e$.

$\tau_e$  is a global quantity associated to the whole basin. A
more detailed description of the transport processes can be
obtained by other suitable Lagrangian quantities such as
residence times \citep{Buffoni1996,Buffoni1997,Falco2000,Orfila2005}. The
particle residence time (RT) is defined as the interval of time
that a fluid particle remains in a region before crossing a
particular boundary. For each fluid particle inside the Bay at
an initial time, we need to compute two times: the forward exit
time, $t_f$, computed as the time needed for a particle to
cross the line delimiting the Bay, taking the forward-in-time
dynamics; and the backward exit time, $t_b$, the same but in
the backward-in-time dynamics. The residence time is defined as
$RT=t_f + t_b$. RTs can be displayed in plots named Lagrangian
Synoptic Maps \citep{Lipphardt2006}, in which the residence
time of each fluid particle is referenced to its initial
position on the grid.

\section{Preliminary Eulerian description}
\label{sec:vel}

A first approach to the transport process in the Bay can be a
description from the Eulerian point of view, by studying
averages of the velocity field. To do this we consider
separately the meridional $v_y$ and zonal $v_x$ components of
the surface flow, and we analyze the time evolution of their
spatial averages.

\begin{figure*}[htb]
\begin{center}
\includegraphics[width=0.55\textwidth, angle=270]{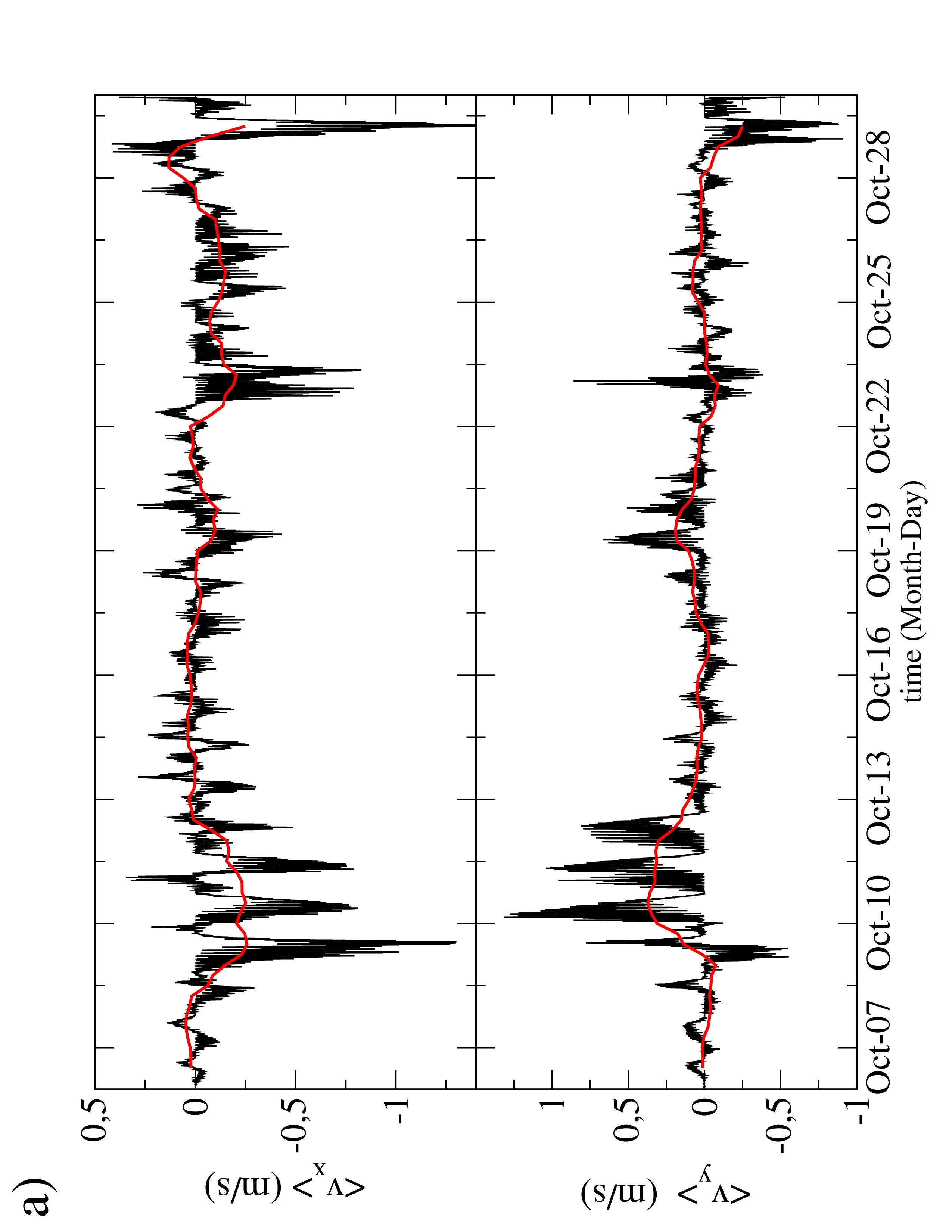}
\includegraphics[width=0.55\textwidth, angle=270]{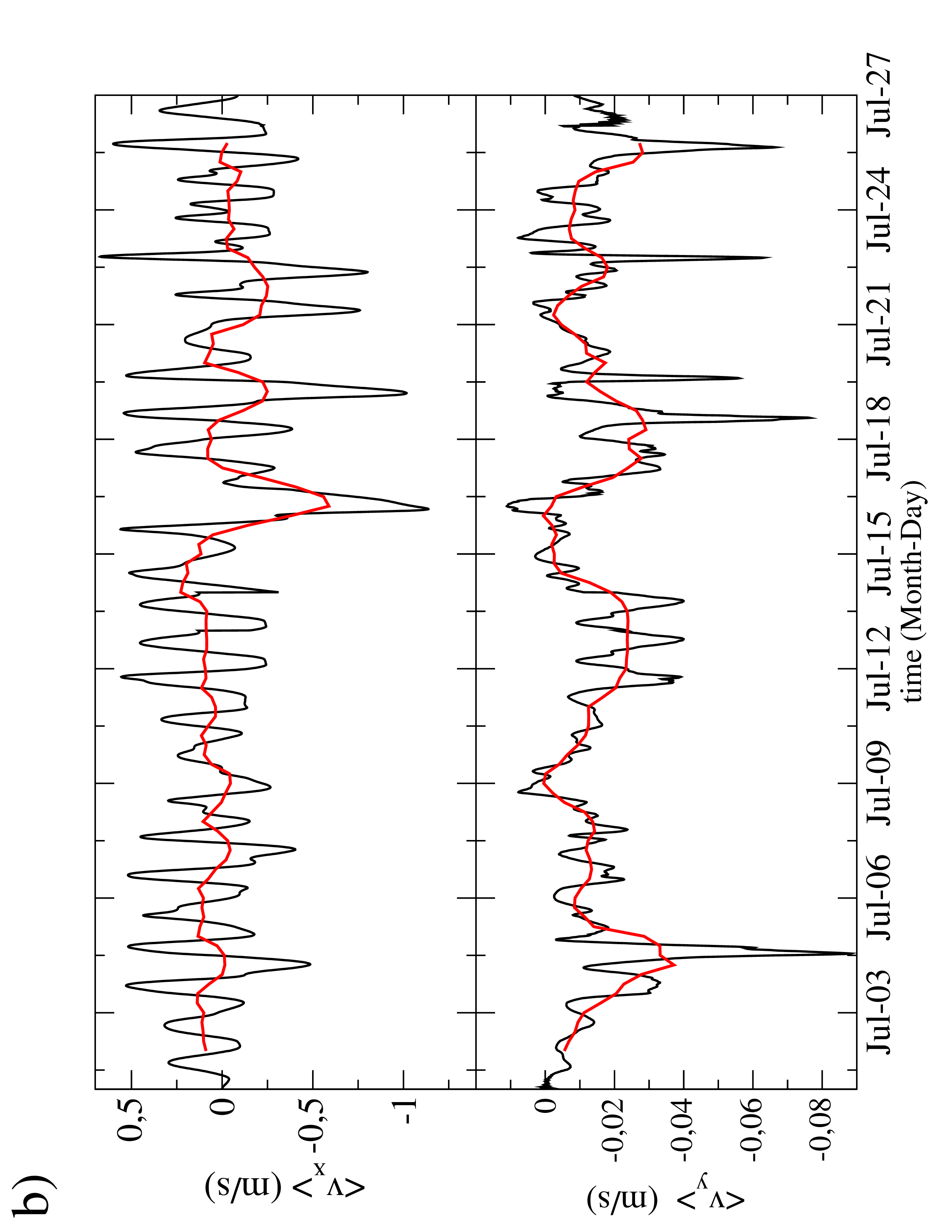}
\end{center}
\caption{a) Complete time series throughout October of the
zonal (top panels) and meridional (bottom panels) of the 
spatial average of the surface velocity
field (black line). The red line is a running daily average. b) the same
as a) but for July. } \label{fig:velo}
\end{figure*}

\begin{figure*}[htb]
\begin{center}
\includegraphics[width=0.73\textwidth, angle=270]{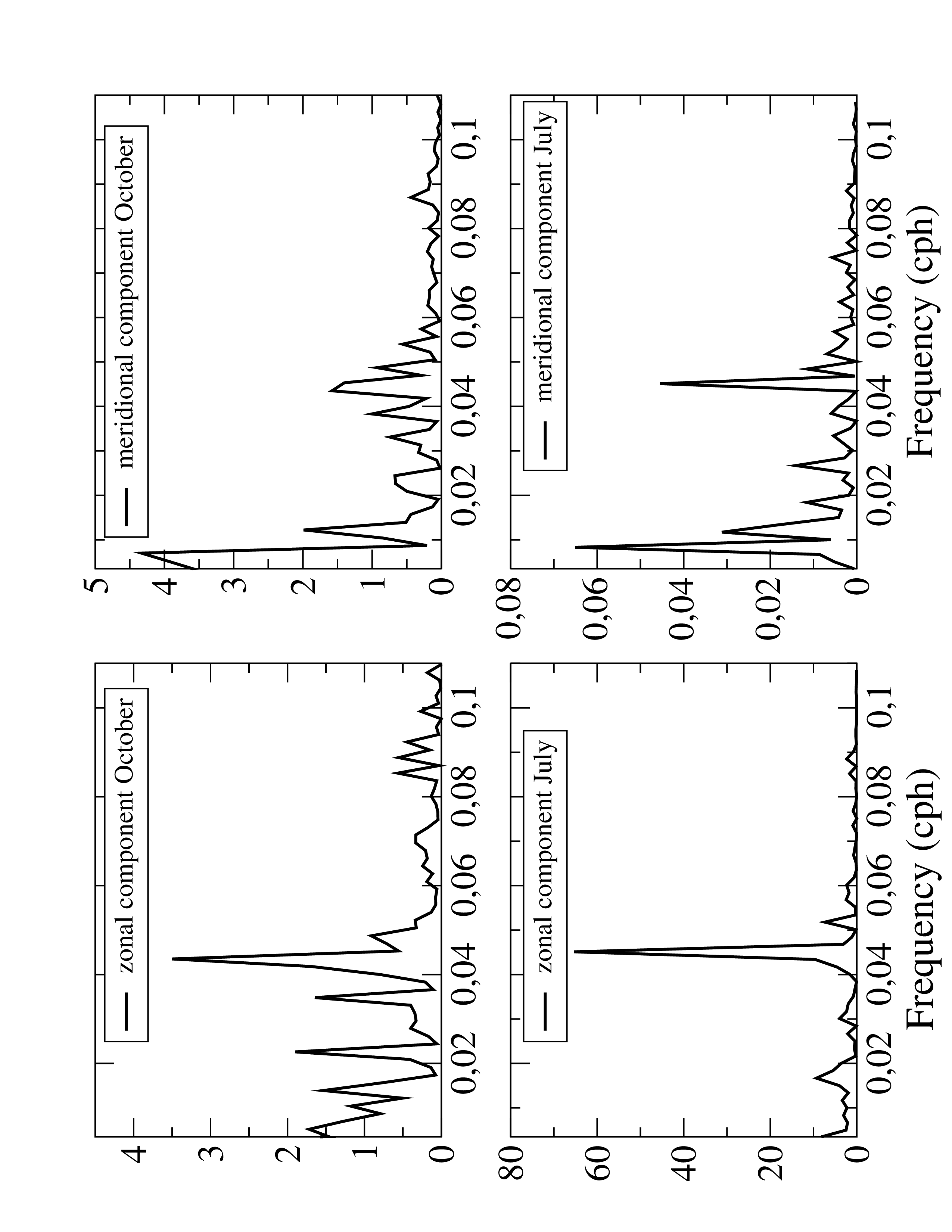}
\end{center}
\caption{Spectra for the zonal (left panels) and meridional component (right panels) 
of the surface velocity field ($m^{2}/s$)in October (top) and July (bottom)} \label{fig:espectro}
\end{figure*}

Figures \ref{fig:velo} a) and b) show the time series of data
taken every 15 min in October and 10 min in July (black
lines), and daily average time series (red lines) of $v_{x}$
and $v_{y}$ for October and July, respectively. The
impact of the more variable and stormy weather in October is
clear in the high frequency variability of the time-series.
During the two months both components of the flow present daily
variability related to the presence of land and sea breezes. In
July the zonal fluctuations are much more noticeable and
regular than the meridional ones, being $\langle v_y \rangle$
very small. We have computed the power spectra for both months 
(see Fig. \ref{fig:espectro}).
In October, in addition to higher power at high frequencies, there are 
also stronger low-frequency fluctuations. From such features in their 
spectra of ADCP-derived velocities \citet{Jordi2011} identified wind-induced 
island trapped waves as the main source of variability in the Bay 
dynamics, in addition to the local wind (essentially sea
breeze). In contrast, the dominant role of sea breeze in 
July is seen as the very strong dominance of the daily 
frequency peak at the July zonal spectrum.

Comparing the velocity components of both months we observe
quantitative differences. The values of $v_{y}$ in the case of
October range from -1.0 to 1.5 $m/s$ (bottom panel of
Fig. \ref{fig:velo} a), while in the case of July, $v_{y}$ is
two orders of magnitude smaller, ranging from -0.1 to
0.02 $m/s$ (bottom panel of Fig. \ref{fig:velo} b). On other
hand, $v_{x}$ are similar during October and July. In October,
$v_{x}$ ranges from -1.5 to 0.5 $m/s$ (top panel of
Fig. \ref{fig:velo} a), the same order of magnitude than the
meridional velocity, resulting in circular motions (clockwise
along the Bay). In July the situation is significantly
different. The zonal velocity ranges from -1.5 to 0.7
$m/s$ (top panel of Fig. \ref{fig:velo} b), much larger than
the meridional velocity, resulting in a flow consisting on
oscillations along the zonal direction. In October the
mean values (and the standard deviations given in parenthesis)
of the time series are $<v_{x}>=-0.0704~(0.1897) m/s$ and
$<v_{y}>=0.0440~(0.1982) m/s$. In July we have
$<v_{x}>=0.0013~(0.3052) m/s$ and $<v_{y}>=-0.0140~(0.0134)
m/s$. The large standard deviation in the zonal velocity in
July is an indicator of the large (breeze induced) daily
fluctuations in this month, but restricted to a single
direction of motion.

\section{Lagrangian Results}
\label{sec:results}

\subsection{Average characterization of stirring}
\label{stirring_mallorca}

We now describe our Lagrangian results. First we compute the
temporal average (over the months of October and July) of the
FSLEs for the surface layer, and for July in the bottom layer.
This calculation helps us to unveil areas of different stirring
and the differences between layers and months.

\begin{figure*}[htb]
\begin{center}
\includegraphics[width=0.99\textwidth, height=0.60\textheight]{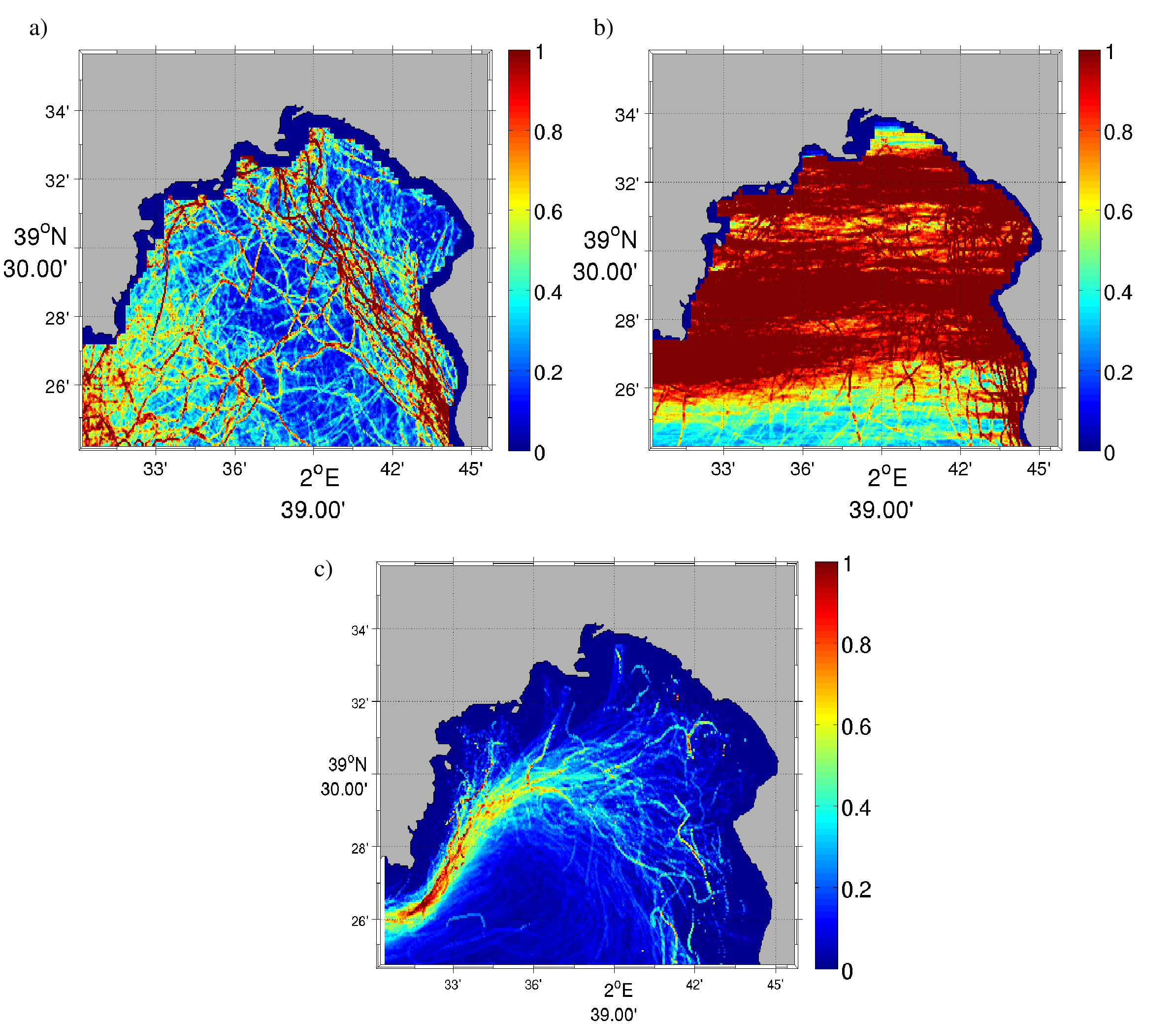}
\end{center}
\caption{Spatial distribution of the time average of 6-hourly FSLEs maps over different months and at
different layers:
a) October at surface layer,
b) July at surface layer,
c) July at bottom layer.
}
\label{fig:promedios_FSLE}
\end{figure*}

The surface computations for the different seasonal months,
October and July (Fig. \ref{fig:promedios_FSLE} a, b) show
different values and spatial distributions of stirring.
We use the same colorbar to compare the stirring in
both months. The Bay of Palma appears to be an area with
important activity. Average FSLE field looks more homogeneous
in July than in October. During October filamental structures
of high values of FSLE are accumulated over the northeast side
of the Bay, forming a linear structure running from north to
south-east which comes from similar structures in the
instantaneous (non-averaged) fields that can act as barriers,
therefore dividing the Bay in two flow regions of qualitatively
different dynamics. The difference in wind regularity and
intensity between these months, and the fact that local and
remote winds are the main drivers of the bay dynamics, explains
the difference in mean stirring distribution among the two
months. The importance of wind will be replaced by bottom
topography when going to the deep layers. The effect of the
terrain topography on stirring is clear in Fig.
\ref{fig:promedios_FSLE} c), where FSLEs are computed at the
deepest layer for July. The high values of time-averaged FSLEs
are located close to a region of high bathymetry gradient,
which seems to act as a barrier along which the flow is
stretched.

\begin{figure*}[htb]
\begin{center}
\includegraphics[width=0.99\textwidth, height=0.55\textheight]{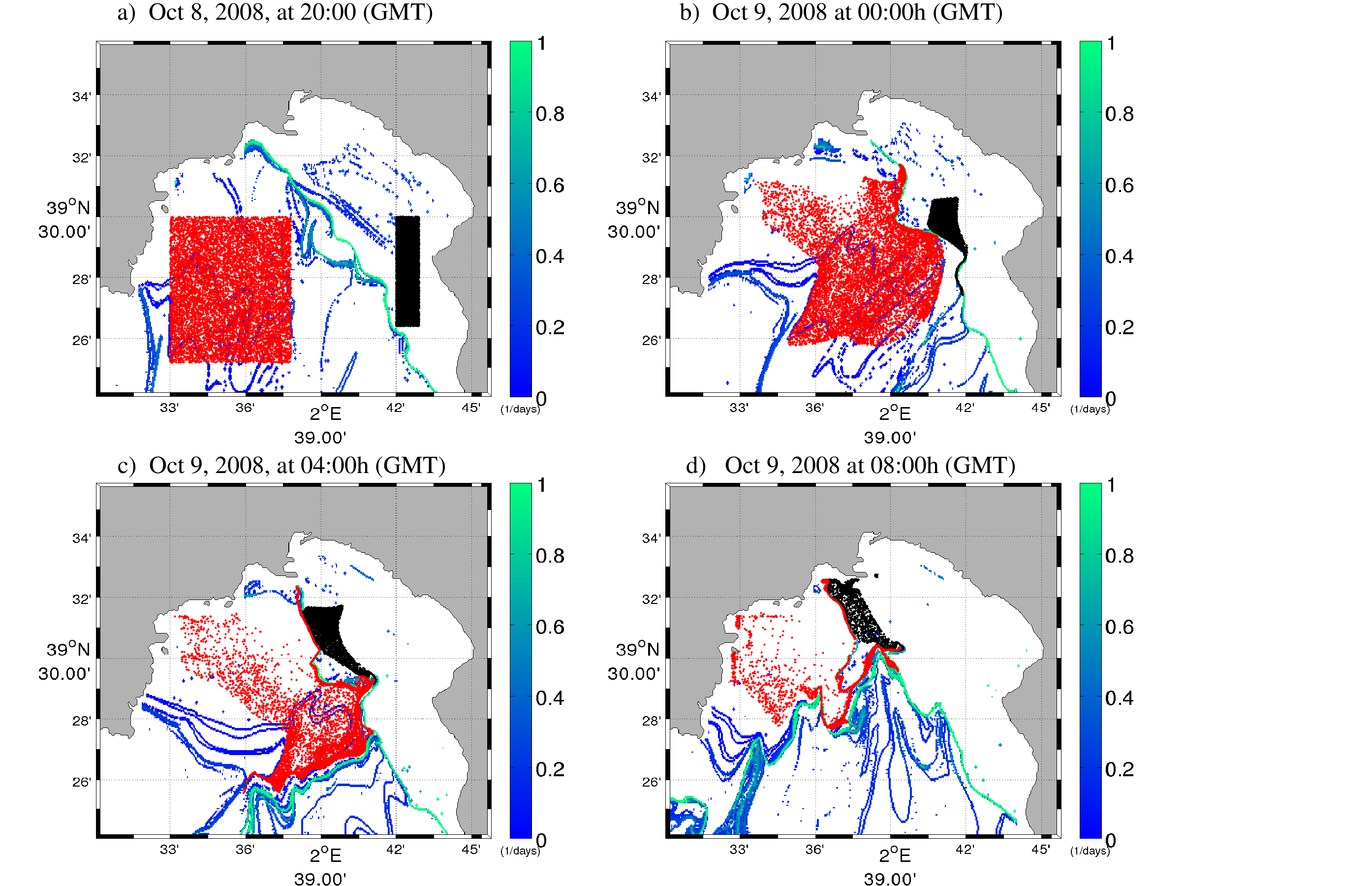}
\end{center}
\caption{Evolution of the locations of two sets of particles in
the Palma Bay during a night of October, superimposed on the
spatial distributions of high values of backward FSLEs. The
colorbars specify FSLE values in days$^{-1}$. Zero FSLE values,
displayed as white, are assigned to locations for which the
particles do not attain the prescribed $\delta_f=10km$
separation after 5-days integration. Note the highest values of
FSLEs (green lines) act as a barrier practically dividing the
Bay in two parts. The two sets of particles are deployed from
both sides of the barrier. (a) Initial conditions of the
particles on October 8 at 20:00 GMT, 2008.; (b) October 9 at
00:00 GMT, 2008 ; (c) October 9 at 04:00 GMT, 2008 ; (d)
October 9 at 08:00 GMT, 2008. Particles marked by black dots
were released in the right side (northeast) of the barrier
while the particles marked with red were released on the left
side of the barrier.  } \label{fig:LCS_oct}
\end{figure*}

\subsection{Coastal LCSs}
\label{sec:LCSs}

The temporal averages computed in Sec. \ref{stirring_mallorca}
give us a rough idea of stirring in the Bay. More detailed
information is obtained by looking at non-averaged quantities,
that may reveal the existence of barriers to transport. Figure
\ref{fig:LCS_oct} shows the location of the high backward FSLE
values (LCSs), appearing as a network of lines, computed at
successive instants of time in October. These temporary
structures can remain for one or more days, as happens in
October, or they can appear in the same location periodically
(not shown). We stress here the appearance of a clear barrier,
from north to south-east, that divides the Bay in two areas
that correlates with the temporal average in Fig.
\ref{fig:promedios_FSLE} a). This barrier appears in almost the
same position in different days, remaining without displacing
too much. To effectively see that it acts as a barrier we have
considered the evolution of virtual particles released at both
sides of the barrier. Red and black particles do not mix and
they tend to spread along the barrier (confirming that, as
expected, it is an attracting line).

In July the situation is rather different. Lines of
high Lyapunov exponents (forward and backwards) are mainly
oriented zonally in the bay (except close to the opening to the
sea), which is also the dominant direction of motion. Thus, it
does not seem that they represent hyperbolic LCS, but rather
lines of intense shear between zonally moving strips.

\begin{figure*}[htb]
\begin{center}
\includegraphics[width=0.61\textwidth, height=0.62\textheight, angle=270]{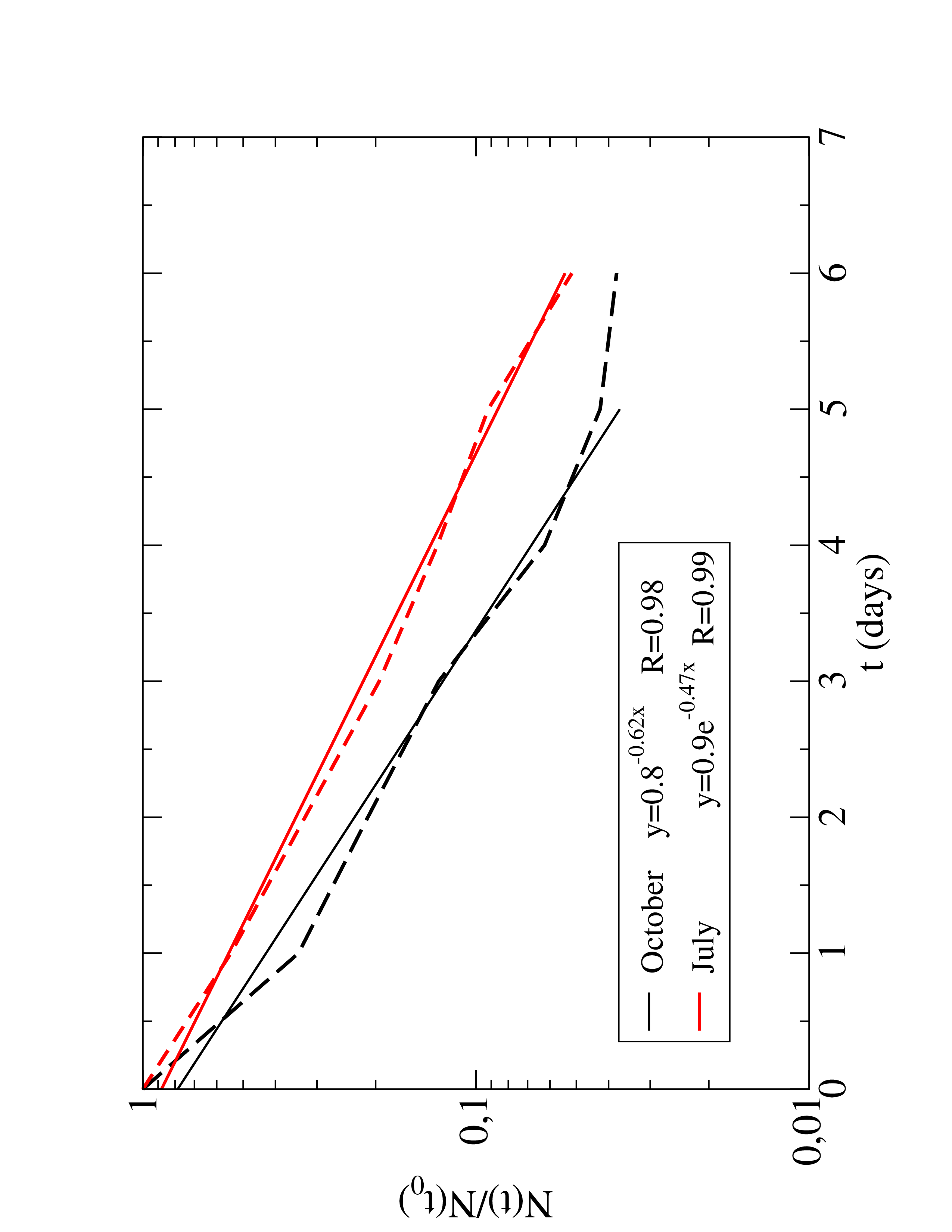}
\end{center}
\caption{Average of 15 subsequent (started at $t_0$ values separated 18 hours)
estimations of $N(t)$, the number of particles remaining in Palma Bay
at least for a lapse of time $t$ after release at $t_0$. Black and red
lines are for surface layer in October and July respectively. Dashed
lines are the measured averages, and the solid lines are the indicated exponential fits.
}
\label{fig:promedios_muestreo}
\end{figure*}

\subsection{Transport between the Bay of Palma and the open sea}
\label{sec:interchange}

In this section we study the surface transport of particles in
and out of the Bay. To have an idea of the time scales involved
in this interchange we proceed by computing the number of
particles remaining in the Bay, $N(t)$, averaged over different
starting times (separated by $18 $hours in order to collect the
information of diurnal and nocturnal signal; this gives us 15
different simulations to be averaged for each month) as a
function of the integration time $t$. A particle is considered
to leave the Bay when crossing the red open-sea boundary in
Fig. \ref{fig:location}, so that particles landing on the coast
are considered not escaped. Fig. \ref{fig:promedios_muestreo}
shows the different average decays for October and July. In
both cases $N(t)$ is reasonably fitted by an exponential in the
considered time-range, thus identifying the escape rates
$\kappa$ = 0.62 and 0.47 $days^{-1}$, respectively. The
corresponding escape times, given by the inverse of the escape
rate, are, respectively, $\tau_e$ = 1.61 and 2.12 $days$.
The relative difference of the escape rates of July with
respect to October, $(\kappa_{October} -
\kappa_{July})/\kappa_{July}$, is 0.32.  Thus the exchange of
fluid particles between the Bay and the open ocean is a
32$\%$ more active in autumn than in summer.

Next we compute synoptic maps of the residence times. As was
indicated in Sec. \ref{sec:methods_ERSLM} the residence time of
the particles throughout the study area is considered as the
sum of the entry time ($t_{b}$) and the escape time ($t_{f}$).
To compute $t_{f}$ and $t_{b}$ particles are initialized every
6 hours in a regular grid of 75$m$ spacing and they are
integrated forward and backward in time during 5 days. We
consider that 5 days is a proper integration time according
with the time scales associated with the coastal processes of
this small Bay, and also owing to the short period of the
available data. In these computations we assign the maximum
possible value of $t_f$ and $t_b$ (5 days) to the fluid
particles that remain in the pre-selected area after the 5 days
of integration.

\begin{figure*}[htb]
\begin{center}
\includegraphics[width=0.91\textwidth, height=0.52\textheight]{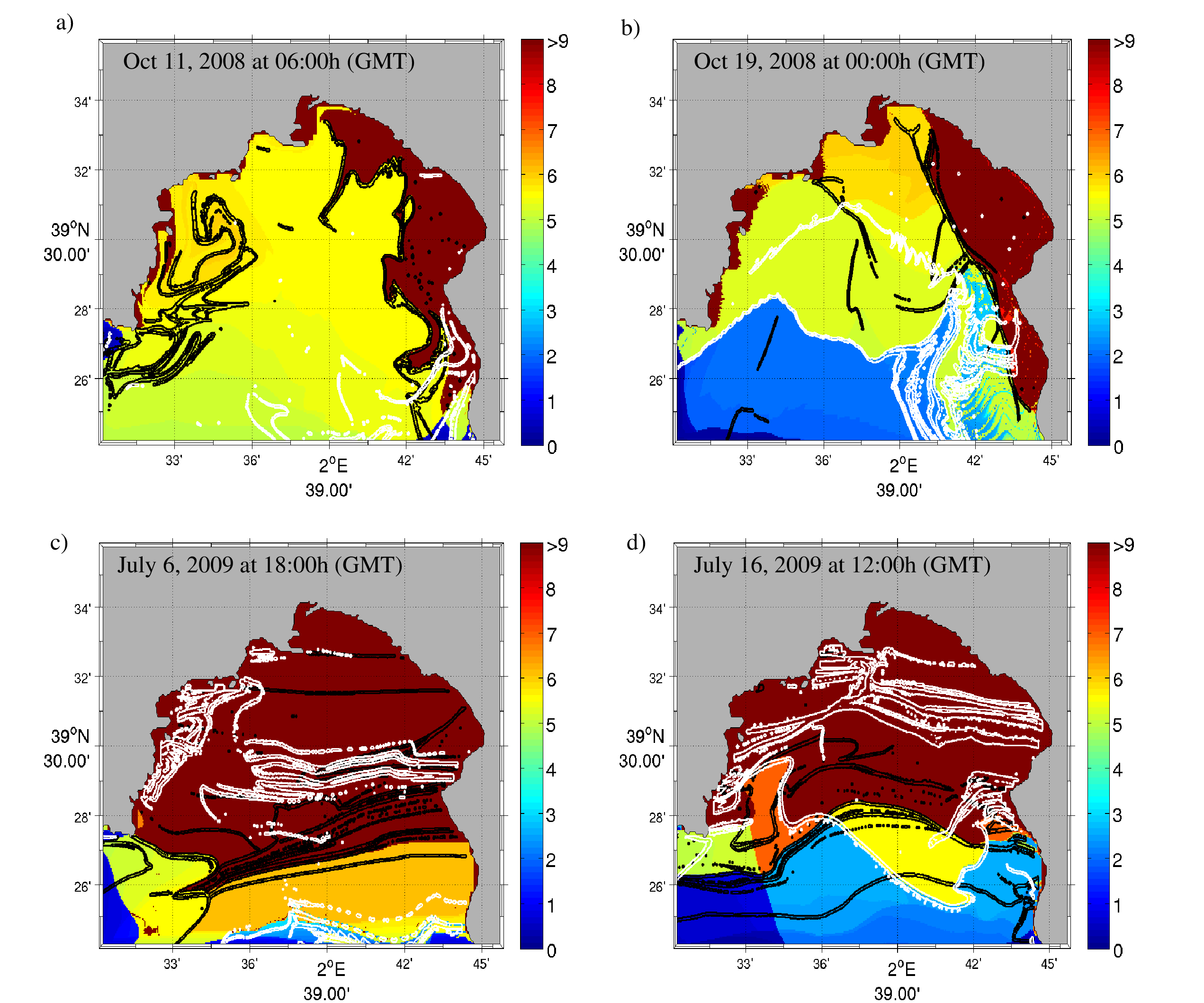}
\end{center}
\caption{Lines are the locations of top values of FSLE (greater
than $0.5days^{-1}$ in October and greater than $1.5days^{-1}$ in July). Backward FSLE
lines are colored in black and forward FSLEs in white. They are
superimposed on spatial distributions of residence times in
Palma Bay for different dates. The colorbars give the
residence times in days. a) and b) correspond to two different
days in October, and c) and d) in July.  }
\label{fig:Snapshots_RT}
\end{figure*}

In Fig. \ref{fig:Snapshots_RT} we color the initial positions
of particles in the Bay attending to the time they transit
through the Bay ($RT=t_f + t_b$) for different days. Initial positions of
particles with short residence times are indicated in blue in
Fig. \ref{fig:Snapshots_RT}. Regions from where particles have
longer residence time (i.e. take more time between entry and
escape) are marked in red/brown.

These maps show that the spatial distribution of residence
times of particles can be complex and time depending,
presenting different patterns at different times. A number of
small structures can be observed, including thin filaments or
small lobes. Comparing both months, one can see differences in
the RT distributions. The most noticeable is the approximate
east-west alignment of the zones of similar RT in July, which
is not seen in October. Also, in October the values of
residence times of the particles are smaller, in agreement with
the global  rate estimations showed before. A common feature is
the southwest region with low values of residence times,
because in this region there are not coastal boundaries and it
is totally open to the ocean.

\begin{figure*}[htb]
\begin{center}
\includegraphics[width=0.90\textwidth, height=0.30\textheight]{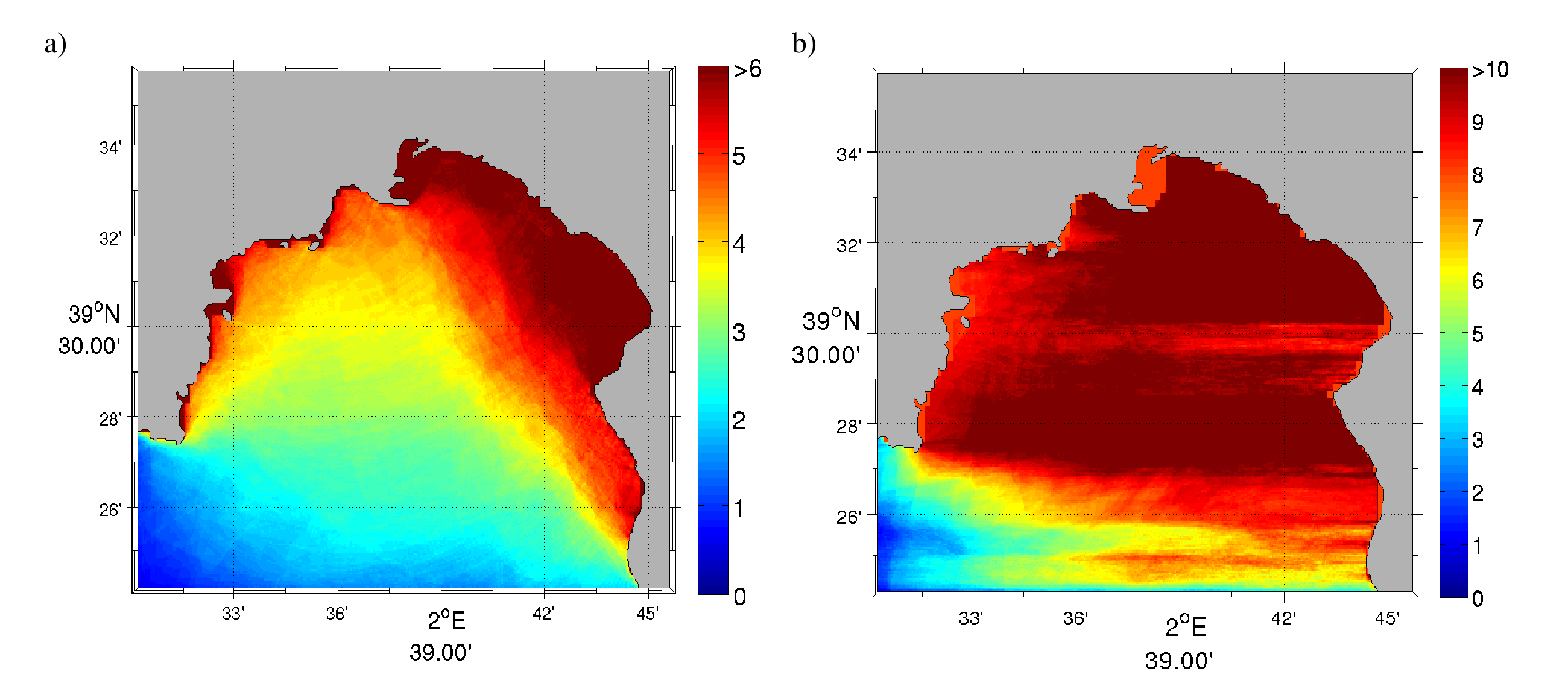}
\end{center}
\caption{Spatial distributions of time averages of 60 snapshots
of 6-hourly RT values collected over 15 days in Palma Bay for
a) October and b) July. The colorbar units are $days$. }
\label{fig:mean_RT}
\end{figure*}

In order to reveal regions with different persistent transport
properties we compute time averages of the spatial
distributions of residence times. We average 6-hourly snapshots
of RT during 15 days (i.e. 60 snapshots) for each month. The
results, plotted in Fig. \ref{fig:mean_RT} a) and b), show the
common features that, in general, the low values of RT are for
particles initiated close to the open ocean, specially in the
southwest part, and high values are for particles started near
the coast, as expected. However, on average, the residence time
is larger in July than in October (3.25 days in July and 1.51
days in October) consistent with the behavior of the
corresponding values of $\tau_e$. Also, in July there is a
clear boundary between the interior of the Bay to the north,
with large average residence times and the open sea to the
south, whereas the boundary between high and low residence
times in October is well inside the Bay, aligned with the
Lagrangian structure identified from the FSLE analysis, as will
be discussed in the next section.

Another feature observed in movies of particle trajectories
(not shown) is that in October fluid particles tend to
circulate mostly clockwise, while in July they are oscillating
along the zonal direction (see Section \ref{sec:vel}). This
difference, arising as discussed in Sect. \ref{sec:vel} from
the different regimes of wind forcing, is likely to be
responsible for most of the different behavior between both
months.

\subsection{Relation between LCSs and RTs}
\label{sec:connetion}

\begin{figure*}[htb]
\begin{center}
\includegraphics[width=0.91\textwidth, height=0.32\textheight]{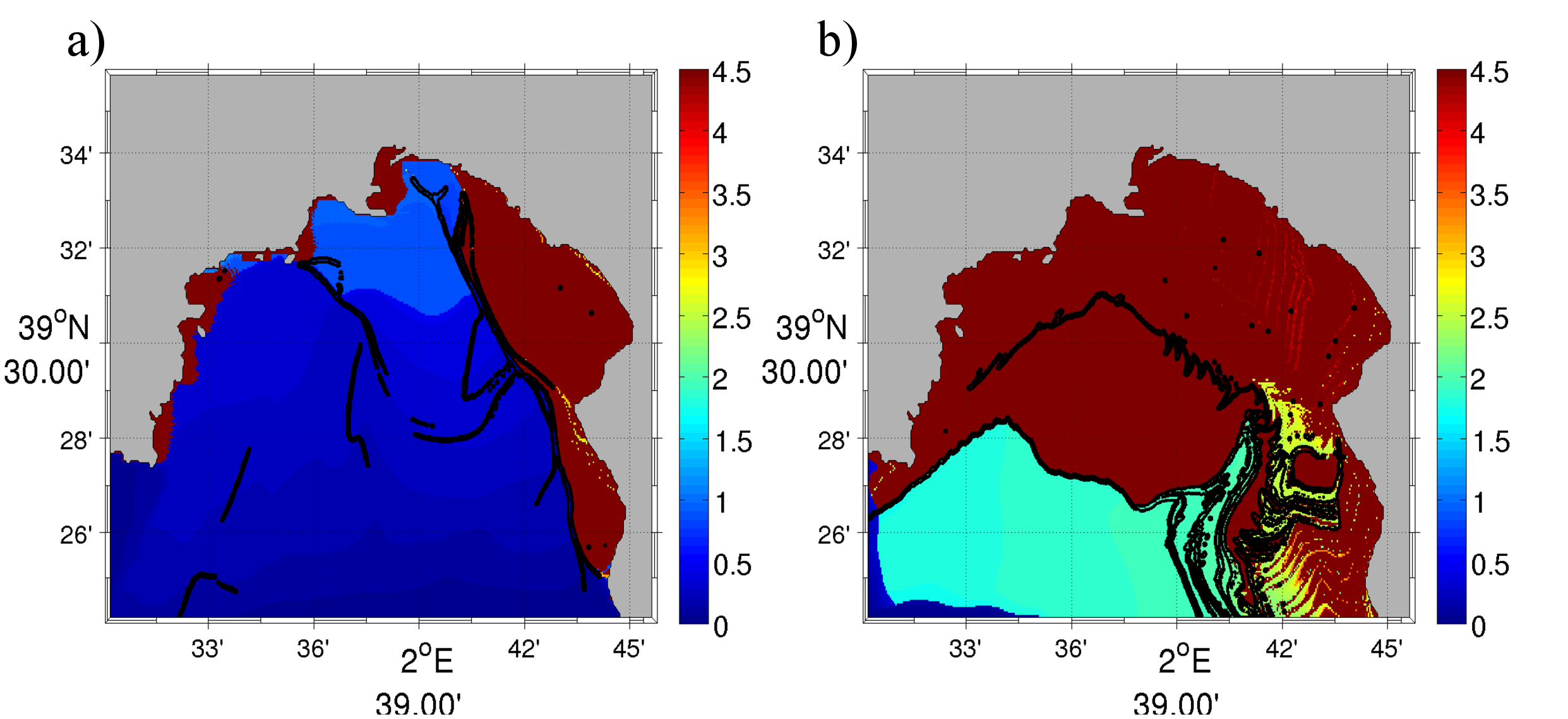}
\end{center}
\caption{Snapshots of top values of FSLE (greater
than $0.5days^{-1}$) plotted over residence 
times. a) is for backward integration in time, and b) is for forward integration. 
Both plots correspond 
to October 19th at 00:00h (GMT) The colorbar units are $days$. }
\label{fig:Snapshots_tb_tf}
\end{figure*}

We now examine the connection between regions of different
residence times with LCSs. To compare RT and FSLE we have
superimposed  in Fig. \ref{fig:Snapshots_RT} the filaments of
high values of forward (white) and backward FSLE (black) values
on the spatial distribution of residence times. The Figure
shows a good correspondence between many structures of
RT and FSLE. Note that RT is the sum of a forward and a
backward exit time, so that some of the strong gradients of RT
will correspond to forward and some others to backwards LCSs.
Fig \ref{fig:Snapshots_tb_tf} shows clearer the correspondence 
between FSLE backwards in time with entry times, and FSLE forward 
with escape times.
At least in October, the LCSs given by high values of FSLE
clearly separate regions with different values of residence
times, confirming the value of the FSLE technique to identify
boundaries between different flow regions and barriers to
transport. There are however some lines of FSLE that
are not associated to gradients of RT, and viceversa. For the
first this happens mainly because we only integrate 5 days
backward and 5 days forward in time, and the time assigned to
the particles that not cross the open-ocean boundary in the 5
days of integration corresponds to the maximum time integration
(5 days). This makes the spatial distribution of the residence
time more homogeneous. We need to integrate the trajectories
longer to unveil more areas with different RT.
In the same way not all abrupt changes in RT are captured by
FSLE lines since we only plot the highest ridges. This
illustrates that both techniques have limitations and that the
complementary use of both could give a rather complete overview
of the geometric structure of flow in marine areas.
\cite{Pattantyus2008} studied the relation between residence
time and FSLE for a wind-forced hydrodynamical model of a
shallow lake. They found that areas with long residence time
visualize the stable manifolds of the so-called chaotic saddle,
a structure controlling the escape properties at long times. In
our case, our integrations are restricted to times too short to
characterize long-time chaotic behavior, but still
there is a good correspondence between the FSLE
structures characterizing attracting or repelling trajectories,
and escape or residence times.


Fig. \ref{fig:mean_RT} a) shows a time average over October of
the spatial distribution of RT, to be compared with the
corresponding average figure (Fig. \ref{fig:promedios_FSLE} a)
for FSLE. It is evident that the region in the north and east
side of the Bay with high values of RT is separated from the
rest by a region of high values of FSLE. This can be explained
by the presence of persistent barriers which do not allow
particles to escape from the northeast side of the Bay, and
thus separating the Bay in regions with different residence
times. In July the situation is different, because the spatial
distribution of FSLE (Fig.\ref{fig:promedios_FSLE} b) and RT
(Fig. \ref{fig:mean_RT} b) is almost homogeneous, with higher
values over the whole area of the Bay, and lower values in the
small region bordering the open ocean. This indicates
that the instantaneous configurations of high FSLE lines (Fig.
\ref{fig:Snapshots_RT}) are not persistent, and that only there
is a persistent large difference between the interior of the
Bay and its opening to the ocean. The predominantly zonal
direction of particle motion in the Bay is consistent with the
orientation of the boundaries between areas of different RTs
in July.

\subsection{Variability of RT and FSLE}
\label{sec:flow}

The differences of residence times and FSLEs in the two
considered months indicate that the dynamics of the flow is
qualitatively different, as anticipated by the
different wind regimes that are the main drivers of the Bay.
Now we analyze the time evolution of their spatial averages.

\begin{figure*}[htb]
\begin{center}
\includegraphics[width=0.63\textwidth, angle=270]{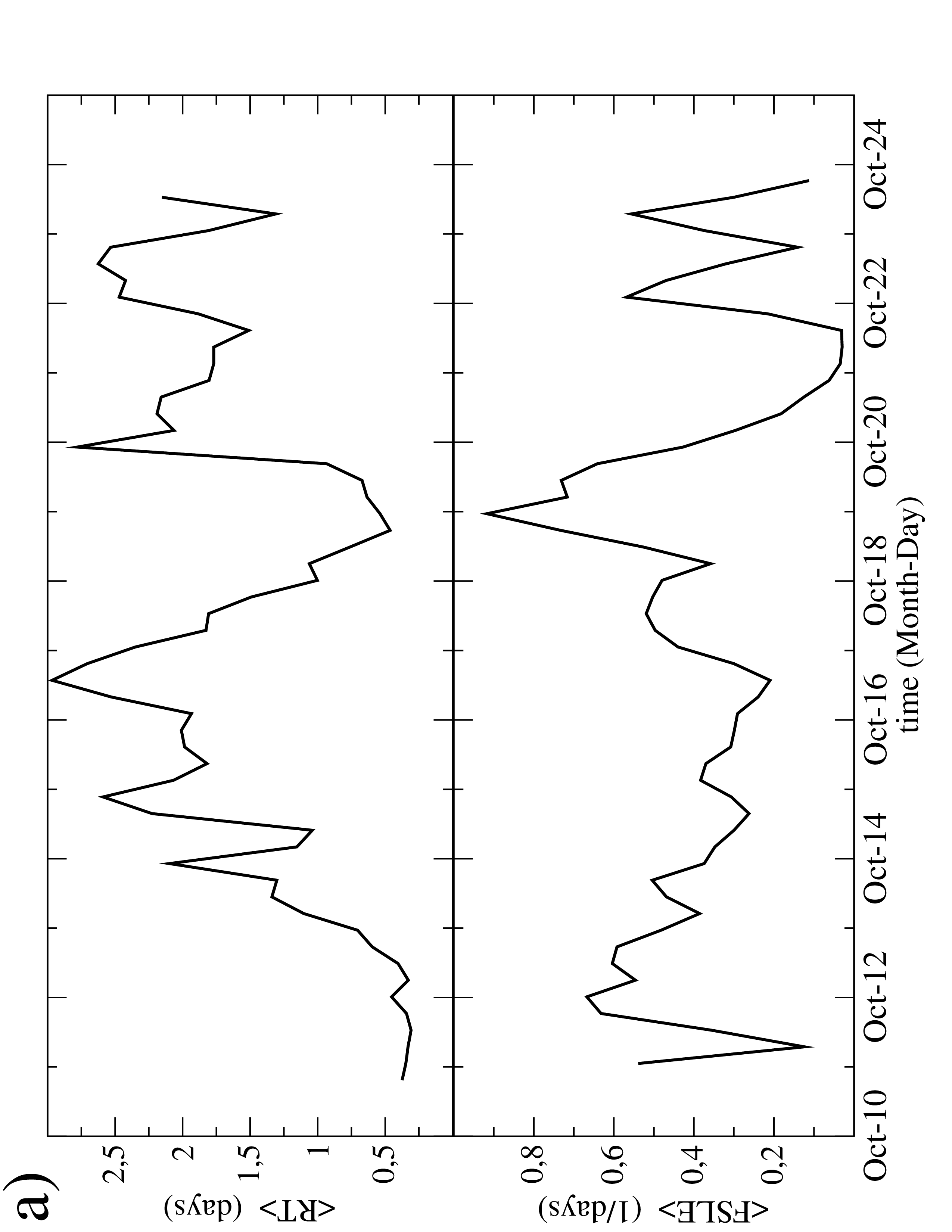}
\includegraphics[width=0.63\textwidth, angle=270]{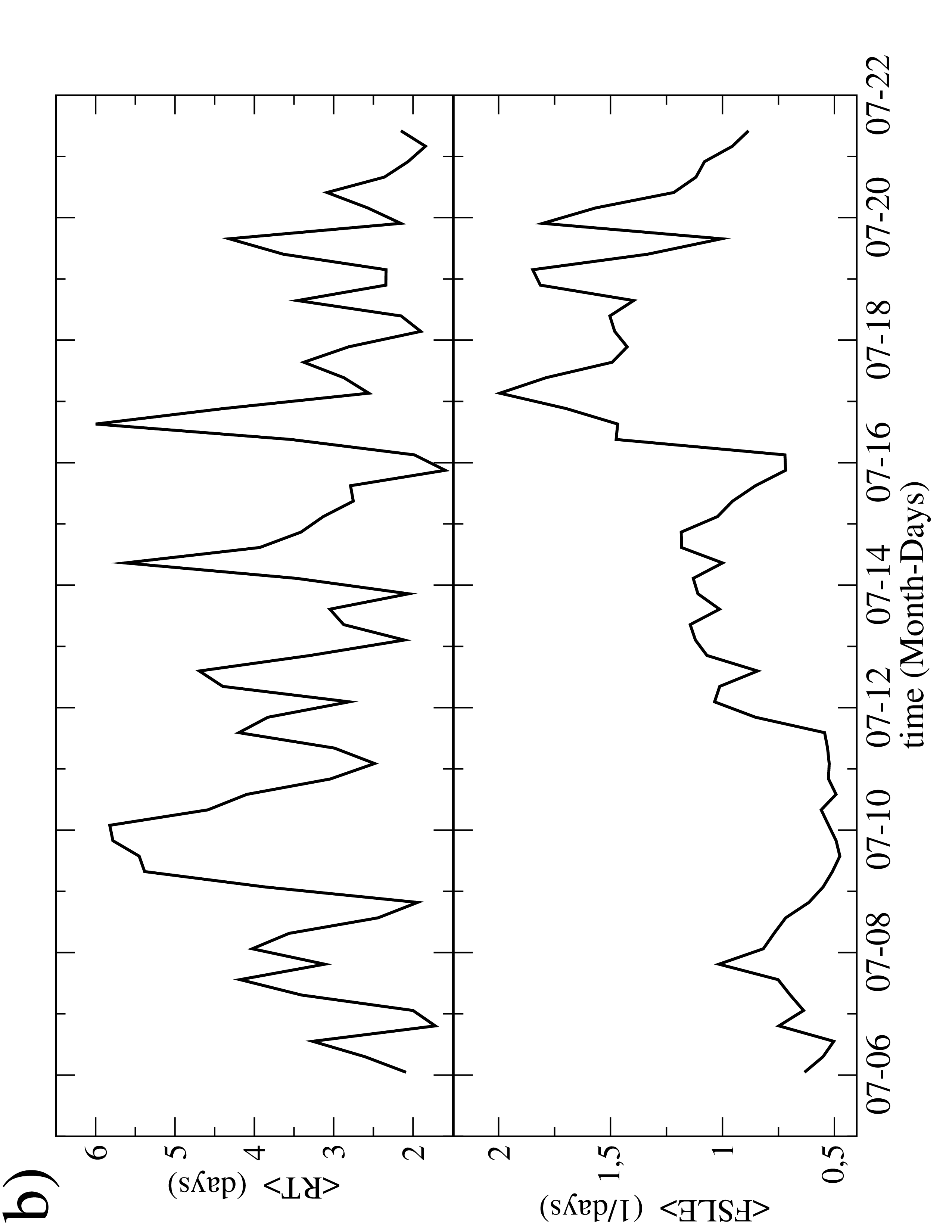}
\end{center}
\caption{a) Time series throughout October of the spatial
average of residence times (top) and spatial average of FSLE
(bottom) for the surface layer of Palma Bay. b) the same that
a) but for July. } \label{fig:FSLE_RT}
\end{figure*}

Figures \ref{fig:FSLE_RT} a) and b) show the time series of the
spatial mean of residence times (top panel) and
backward FSLE (bottom panel) for October and July,
respectively.

Comparison between time evolution of spatial average of the RT
for the different months confirms, again, that particles tend
to stay longer times in the Bay during July than in October.
The values of RT vary approximately from 0.25 days to 3 days in
October (Fig. \ref{fig:FSLE_RT} a, top), and from 2 days to 6
days in July (Fig. \ref{fig:FSLE_RT} b, top). The same happens
with FSLE, higher values correspond to July and lower ones to
October. Diurnal fluctuations, likely related to the effect of
the sea breeze, are evident in RT and FSLE for both months. In
October there are some large fluctuations of low frequency in
RT, probably induced by the variability of remote forcing
winds. On the other hand, during October, minima of RT
correspond to maxima of FSLE, and maxima of RT correspond to
minima of FSLE. In July, the relationship between FSLE and RT
is looser and only observed in the high-frequency fluctuations.
To be more quantitative, Pearson correlation
coefficient between RT and FSLE time series is -0.526 in
October, but just -0.223 in July. This difference in behavior
probably arises from the fact that high values of FSLE
determine clear and well-defined barriers to transport only in
the case of October (see Sect. \ref{sec:interchange} and
\ref{sec:connetion}). In July lines of high FSLE remain
nearly zonal and parallel to dominant particle direction of
motion. Thus, as commented in Sect. \ref{sec:LCSs}, they
probably represent regions of high zonal shear everywhere in
the Bay. Their high or low values indicate large or small
differences in East-West velocities but by themselves they do
not imply stronger or weaker escape towards the South.

\begin{figure*}[htb]
\begin{center}
\includegraphics[width=0.95\textwidth, height=0.32\textheight]{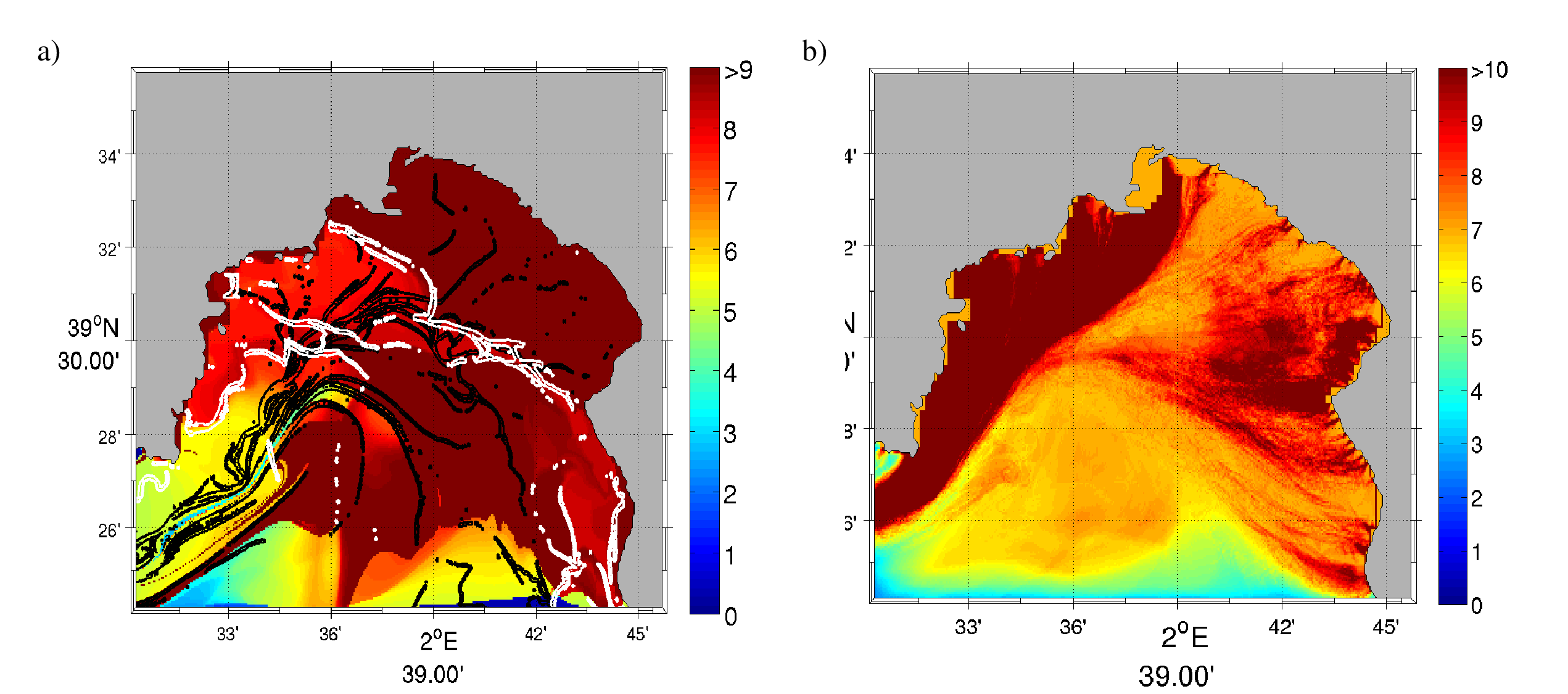}
\end{center}
\caption{a) Snapshot of spatial distributions of residence
times at the bottom layer in Palma Bay corresponding to July
17, 2009 at 18:00h (GMT). Lines are the locations of top values of FSLE (greater
than $0.3days^{-1}$). Backward FSLE
lines are colored in black and forward FSLEs in white.  b) spatial distribution of time
average of 60 snapshots of 6-hourly maps of RT collected over
15 days in July at the bottom layer.}
\label{fig:mean_RT_bottom}
\end{figure*}

\subsection{Transport at the bottom layer}

In this subsection we compare the main Lagrangian
characteristics at the bottom layer, not driven directly by
wind, with those at the surface. In Fig.
\ref{fig:mean_RT_bottom} a) we show an instantaneous map of the
residence times in the bottom layer one day of July, overlayed
with lines of high FSLE values. Again, the spatial distribution
is inhomogeneous, and we find high values of RT over all the
Bay except very close to the ocean. The correlation of RT
values with FSLE lines is weaker than in the upper layer, but
still we see that the relatively lower values of RT in the
western part of the Bay at that particular day appear bounded
by backwards FSLE lines, indicating a temporal escape route of
particles in that region towards the southwest. The spatial
distribution of the time average of RT plotted in Fig.
\ref{fig:mean_RT_bottom} b) shows that highest average values
of RT are concentrated in the northwestern region of the Bay.
Fig. \ref{fig:promedios_FSLE} c) displays high values of FSLE
located precisely in the same region where the RT qualitatively
change to high values. This suggests the presence of persistent
barriers that separate this southeastern region from the rest
in this bottom layer. The formation of these persistent LCSs
is associated to the gradient of the bathymetry (see
Fig. \ref{fig:location}).

\begin{figure*}[htb]
\begin{center}
\includegraphics[width=0.63\textwidth, angle=270]{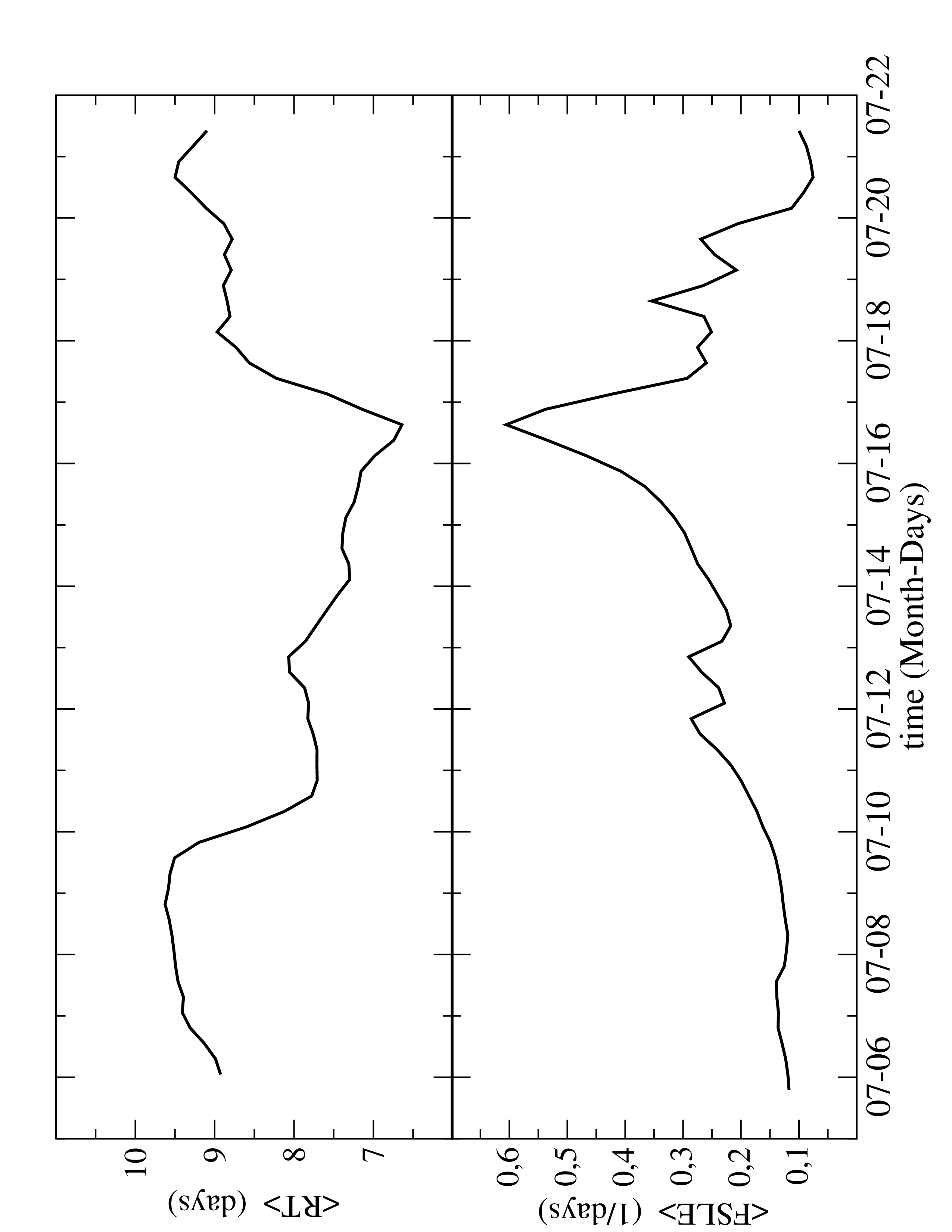}
\end{center}
\caption{Time series throughout July of (top) spatial average of residence times
and (bottom) spatial average of FSLE computed for the bottom layer.
}
\label{fig:FSLE_RT_bottom}
\end{figure*}

The time evolution of the spatial average of RT and FSLE are
plotted in top and bottom panel of Fig.
\ref{fig:FSLE_RT_bottom} respectively. Contrarily to the
surface, in the bottom layer the diurnal fluctuations in the
time series of RT disappear, showing that the flow at this
depth is not directly influenced by breeze. The RT values are
larger than in the surface, and therefore the interchange
between the ocean and the Bay is less intense at bottom layers.
This is a consequence of the slowness of the flow produced by
the absence of direct wind forcing at deepest levels. There is
a strong negative correlation ($r=-0.803$) between stirring and
residence times: when the flow is more dispersive the particles
transit during less time over the Bay, so that maxima of RT
correspond to minima of FSLE and vice-versa.

\section{Conclusions}
\label{sec:conclusion}

Properties of coastal transport in the Bay of Palma, which is a
small semi-enclosed region of the Island of Mallorca, were
studied in a Lagrangian framework, by using model velocity data
at high resolution. We have applied two complementary
Lagrangian methods (FSLEs and RT) to analyze the small scales
of these coastal currents. LCSs have been detected as
high ridges of FSLE, and virtual experiments with
particle trajectories have shown that these structures really
act as barriers in most cases, organizing the coastal
flow. Global and average aspects of the transport in different
seasonal months show that, in the period studied, in autumn
there is more exchange between the Bay and the open
ocean than in summer. This arises from the different
wind regimes in both months, that during July induce a flow
that restricts motion of the coastal marine surface to the zonal
direction, preventing the flow to enter or escape toward the
open ocean. The transport of particles at the deepest layer is
less active than at the surface and not directly driven by wind,
but influenced by the bottom topography. Regions with
different values of RT are generally separated by ridges of
FSLE, proving the fact that FSLE separate regions of
qualitatively different dynamics also in small coastal regions.
Thus, we think that these Lagrangian quantities can be used as
key variables able to determine the dynamics and health of
other bays or estuaries, particularly in relation with human
activities. Future improvements include the adaptation of these
methods to three-dimensional spaces and capture
three-dimensional effects, such as upwelling and downwelling in
coastal areas, and analyzing longer periods of time.

%
%

\section*{Acknowledgments}

I.H-C, C.L and E.H-G acknowledge support from MICINN and FEDER
through projects FISICOS (FIS2007-60327), INTENSE@COSYP
(FIS2012-30634), and ESCOLA (CTM2012-39025-C02-01).
AO acknowledges financial support from MICINN and EU-Med Programme
through projects BUS2 (CGL2011-22964) and TOSCA (2GMED-09-425).
We acknowledge the ICMAT Severo Ochoa project (SEV-2011-0087)
for the funding of the publication charges of
this article.


\begin{thebibliography}{47}
\providecommand{\natexlab}[1]{#1} \providecommand{\url}[1]{{\tt
#1}} \providecommand{\urlprefix}{URL } \expandafter\ifx\csname
urlstyle\endcsname\relax
  \providecommand{\doi}[1]{doi:\discretionary{}{}{}#1}\else
  \providecommand{\doi}{doi:\discretionary{}{}{}\begingroup
  \urlstyle{rm}\Url}\fi

\bibitem[{Antonov et~al.(2006)Antonov, Locarnini, Boyer,
    Mishonov, and
  Garcia}]{Antonov2006}
Antonov, J.~I., Locarnini, R.~A., Boyer, T.~P., Mishonov,
A.~V., and Garcia,
  H.~E.: World Ocean Atlas 2005, Volume 2: Salinity. S. Levitus, Ed. NOAA Atlas
  NESDIS 62, U.S. Government Printing Office, Washington, D.C, 2006.

\bibitem[{Aurell et~al.(1997)Aurell, Boffetta, Crisanti,
    Paladin, and
  Vulpiani}]{Aurell1997}
Aurell, E., Boffetta, G., Crisanti, A., Paladin, G., and
Vulpiani, A.:
  Predictability in the large: an extension of the {Lyapunov} exponent, J.
  Phys. A, 30, 1--26, 1997.

\bibitem[{Boffetta et~al.(2001)Boffetta, Lacorata, Redaelli,
    and
  Vulpiani}]{Boffetta2001}
Boffetta, G., Lacorata, G., Redaelli, G., and Vulpiani, A.:
Detecting barriers
  to transport: a review of different techniques, Physica D, 159, 58--70, 2001.

\bibitem[{Buffoni et~al.(1996)Buffoni, Cappelletti, and
    Cupini}]{Buffoni1996} Buffoni, G., Cappelletti, A., and
    Cupini, E.: Advection-diffusion processes and
  residence times in semi-enclosed marine basins, Int. J. Numer. Methods
  Fluids, 22, 1--23, 1996.

\bibitem[{Buffoni et~al.(1997)Buffoni, Falco, Griffa, and
  Zambianchi}]{Buffoni1997}
Buffoni, G., Falco, P., Griffa, A., and Zambianchi, E.:
Dispersion processes
  and residence times in a semi-enclosed basin with recirculating gyres: an
  application to the {T}yrrhenian {S}ea, J. Geophys. Res, 102(C8),
  18\,699--18\,713, 1997.

\bibitem[{Calil and Richards(2010)}]{Calil2010} Calil, P. and
    Richards, K.: Transient upwelling hot spots in the
    oligotrophic
  {N}orth {P}acific, J. Geophys. Res, 115, C02\,003, 2010.

\bibitem[{Cencini et~al.(2010)Cencini, Cecconi, and
    Vulpiani}]{libroangelo} Cencini, M., Cecconi, F., and
    Vulpiani: Chaos: {F}rom simple models to complex
  systems, Series on Advances in Statistical Mechanics, World Scientific,
  Singapore, 2010.

\bibitem[{Dobricic et~al.(2007)Dobricic, Pinardi, Adani,
    Tonani, Fratianni,
  Bonazzi, and Fernandez}]{Dobricic2007}
Dobricic, S., Pinardi, N., Adani, M., Tonani, M., Fratianni,
C., Bonazzi, A.,
  and Fernandez, V.: Daily oceanographic analyses by Mediterranean Forecasting
  System at the basin scale, Ocean Science, 3, 149--157, 2007.

\bibitem[{d'Ovidio et~al.(2004)d'Ovidio, Fern\'andez,
    Hern\'andez-Garc\'ia, and
  L\'opez}]{dOvidio2004}
d'Ovidio, F., Fern\'andez, V., Hern\'andez-Garc\'ia, E., and
L\'opez, C.:
  Mixing structures in the {M}editerranean sea from Finite-Size {L}yapunov
  Exponents, Geophys. Res. Lett., 31, L17\,203,
  2004.

\bibitem[{d'Ovidio et~al.(2009)d'Ovidio, Isern-Fontanet,
    L\'opez,
  Hern\'andez-Garc\'{\i}a, and Garc\'{\i}a-Ladona}]{dOvidio2009}
d'Ovidio, F., Isern-Fontanet, J., L\'opez, C.,
Hern\'andez-Garc\'{\i}a, E., and
  Garc\'{\i}a-Ladona, E.: Comparison between {E}ulerian diagnostics and
  {F}inite-{S}ize {L}yapunov {E}xponents computed from Altimetry in the
  {A}lgerian basin, Deep-Sea Res. I, 56, 15--31, 2009.

\bibitem[{Falco et~al.(2000)Falco, Griffa, and
    Poulain}]{Falco2000} Falco, P., Griffa, A., and Poulain,
    P.~M.: Transport properties in the
  {A}driatic {S}ea as deduced from drifter data, J. Phys. Oceanogr., 30(8),
  2055--2071, 2000.

\bibitem[{Fiorentino et~al.(2012)Fiorentino, Olascoaga,
    Reniers, Feng,
  Beron-Vera, and MacMahan}]{Fiorentino2012}
Fiorentino, L.~A., Olascoaga, M.~J., Reniers, A., Feng, Z.,
Beron-Vera, F., and
  MacMahan, J.~H.: Using {L}agrangian {C}oherent {S}tructures to understand
  coastal water quality, Continental Shelf Research, 47, 145--149, 2012.

\bibitem[{Galan et~al.(2012)Galan, Orfila, Simarro,
    Hern\'{a}ndez-Carrasco, and
  L\'{o}pez}]{Galan2012}
Galan, A., Orfila, A., Simarro, G., Hern\'{a}ndez-Carrasco, I.,
and L\'{o}pez,
  C.: Wave mixing rise inferred from Lyapunov exponents, Environmental Fluid
  Mechanics, 12, 291--300, 2012.

\bibitem[{Gildor et~al.(2009)Gildor, Fredj, Steinbuck, and
  Monismith}]{Gildor2009}
Gildor, H., Fredj, E., Steinbuck, J., and Monismith, S.:
Evidence for
  {S}ubmesoscale {B}arriers to {H}orizontal {M}ixing in the {O}cean from
  {C}urrent {M}eauserements and {A}erial {P}hotographs, Journal od Physical
  Oceanography, 39, 1975--1983,  2009.

\bibitem[{Haidvogel et~al.(2000)Haidvogel, Blanton, Kindle, and
  Lynch}]{Haidvogel2000}
Haidvogel, D., Blanton, J., Kindle, J., and Lynch, D.: Coastal
{O}cean
  {M}odeling: {P}rocesses and {R}eal-{T}ime {S}ystems, Oceanography, 13(1),
  35--46, 2000.

\bibitem[{Haller(2011)}]{Haller2011a} Haller, G.: A variational
    theory of hyperbolic {L}agrangian Coherent
  Structures, Physica D, 240, 574 -- 598, 2011.
  Erratum and addendum: Physica D 241, 439--441, 2012.

\bibitem[{Haller and Beron-Vera(2012)}]{Haller2012} Haller, G.
    and Beron-Vera, F.~J.: Geodesic theory of transport
    barriers in
  two-dimensional flows, Physica D: Nonlinear Phenomena, 241, 1680 -- 1702, 2012.

\bibitem[{Haller and Yuan(2000)}]{Haller2000b} Haller, G. and
    Yuan, G.: Lagrangian coherent structures and mixing in
  two-dimensional turbulence, Physica D, 147, 352--370, 2000.

\bibitem[{Haza et~al.(2010)Haza, \"Ozg\"okmen, Griffa, Molcard,
    Poulain, and
  Peggion}]{Haza2010}
Haza, A.~C., \"Ozg\"okmen, T.~M., Griffa, A., Molcard, A.,
Poulain, P.-M., and
  Peggion, G.: Transport properties in small-scale coastal flows: relative
  dispersion from {VHF} radar measurements in the {Gulf of La Spezia}, Ocean
  Dynamics, 60, 861--882, 2010.

\bibitem[{Hern\'{a}ndez-Carrasco
    et~al.(2011)Hern\'{a}ndez-Carrasco, L\'{o}pez,
  Hern\'{a}ndez-Garc\'{\i}a, and Turiel}]{HernandezCarrasco2011}
Hern\'{a}ndez-Carrasco, I., L\'{o}pez, C.,
Hern\'{a}ndez-Garc\'{\i}a, E., and
  Turiel, A.: How reliable are finite-size {L}yapunov exponents for the
  assesment of ocean dynamics?, Ocean Modelling, 36(3-4), 208--218, 2011.

\bibitem[{Huhn et~al.(2012)Huhn, von Kameke, Allen-Perkins,
    Montero, Venancio,
  and P{\'e}rez-Mu{\~n}uzuri}]{Huhn2012}
Huhn, F., von Kameke, A., Allen-Perkins, S., Montero, P.,
Venancio, A., and
  P{\'e}rez-Mu{\~n}uzuri, V.: Horizontal {L}agrangian transport in a
  tidal-driven estuary --Transport barriers attached to prominent coastal
  boundaries, Continental Shelf Research, 39-40, 1--13, 2012.

\bibitem[{Jordi et~al.(2009)Jordi, Basterretxea, and
    Wang}]{Jordi2009} Jordi, A., Basterretxea, G., and Wang,
    D.-P.: Evidence of sediment resuspension
  by island trapped waves, Geophys. Res. Lett., 36, 2009.

\bibitem[{Jordi et~al.(2011)Jordi, Basterretxea, and
    Wang}]{Jordi2011} Jordi, A., Basterretxea, G., and Wang,
    D.-P.: Local versus remote wind effects
  on the coastal circulation of a microtidal bay in the {M}editerranean {S}ea,
  Journal of Marine Systems, 88, 312--322, 2011.

\bibitem[{Joseph and Legras(2002)}]{Joseph2002} Joseph, B. and
    Legras, B.: Relation between {K}inematic {B}oundaries,
  {S}tirring, and {B}arriers for the {A}ntartic {P}olar {V}ortex, J. Atm. Sci.,
  59, 1198--1212, 2002.

\bibitem[{Koh and Legras(2002)}]{Koh2002} Koh, T. and Legras,
    B.: Hyperbolic lines and the stratospheric {P}olar vortex,
  Chaos, 12, 382--394, 2002.

\bibitem[{Lai and Tel(2011)}]{Lai2011} Lai, Y. and Tel, T.:
    Transient Chaos: Complex Dynamics on Finite-Time Scales,
  Springer, 2011.

\bibitem[{Lehan et~al.(2007)Lehan, d'Ovidio, L\'evy, and
    Heyfetz}]{Lehan2007} Lehan, Y., d'Ovidio, F., L\'evy, M.,
    and Heyfetz, E.: Stirring of the
  {N}ortheast {A}tlantic spring bloom: A {L}agrangian analysis based on
  multisatellite data, J. Geophys. Res., 112, C08\,005, 2007.

\bibitem[{Lekien et~al.(2005)Lekien, Coulliette, Mariano, Ryan,
    Shay, Haller,
  and Marsden}]{Lekien2005}
Lekien, F., Coulliette, C., Mariano, A.~J., Ryan, E.~H., Shay,
L.~K., Haller,
  G., and Marsden, J.: Pollution release tied to invariant manifolds: {A} Case
  study for the coast of {F}lorida, Physica D, 210, 1--20, 2005.

\bibitem[{Lipphardt et~al.(2006)Lipphardt, Jr., Small, Kirwan,
    Jr., Wiggins,
  Ide, Grosch, and Paduan}]{Lipphardt2006}
Lipphardt, B., Jr., Small, D., Kirwan, A., Jr., Wiggins, S.,
Ide, K., Grosch,
  C., and Paduan, J.: Synoptic Lagrangian maps: Aplication to surface transport
  in {M}onterey {B}ay, Journal of Marine Research, 64, 221--247, 2006.

\bibitem[{Locarnini et~al.(2006)Locarnini, Mishonov, Antonov,
    Boyer, and
  Garcia}]{Locarnini2006}
Locarnini, R.~A., Mishonov, A.~V., Antonov, J.~I., Boyer,
T.~P., and Garcia,
  H.~E.: World Ocean Atlas 2005, Volume 1: Temperature. S. Levitus, Ed. NOAA
  Atlas NESDIS 61, U.S. Government Printing Office, Washington, D.C, 2006.

\bibitem[{Mancho et~al.(2006)Mancho, Small, and
    Wiggins}]{Mancho2006b} Mancho, A.~M., Small, D., and
    Wiggins, S.: A tutorial on dynamical systems
  concepts applied to Lagrangian transport in oceanic flows defined as finite
  time data sets: Theoretical and computational issues, Physics Reports, 437,
  55 -- 124, 2006.

\bibitem[{Mendoza and Mancho(2010)}]{Mendoza2010} Mendoza, C.
    and Mancho, A.~M.: Hidden Geometry of Ocean Flows, Phys.
    Rev.
  Lett., 105, 038\,501, 2010.

\bibitem[{Mezi\'c et~al.(2010)Mezi\'c, Loire, Vladimir, Fonoberov, and
  Hogan}]{Mezic2010}
Mezi\'c, I., Loire, S., Vladimir, A., Fonoberov, A., and Hogan, P.: A {N}ew
  {M}ixing {D}iagnostic and {G}ulf {O}il {S}pill {M}ovement, Science,
  330(6003), 486--489, 2010.

\bibitem[{Nencioli et~al.(2011)Nencioli, d'Ovidio, Doglioli, and
  Petrenko}]{Nencioli2011}
Nencioli, F., d'Ovidio, F., Doglioli, A., and Petrenko, A.: Surface coastal
  circulation patterns by in-situ detection of {L}agrangian {C}oherent
  {S}tructures, Geophys. Res. Lett., 38, L17\,604, 2011.  
  
  
\bibitem[{Oddo et~al.(2009)Oddo, Adani, Pinardi, Fratianni,
    Tonani, and
  Pettenuzzo}]{Oddo2009}
Oddo, P., Adani, M., Pinardi, N., Fratianni, C., Tonani, M.,
and Pettenuzzo,
  D.: A nested Atlantic-Mediterranean Sea general circulation model for
  operational forecasting, Ocean Science, 5, 461--473, 2009.

\bibitem[{Ohlmann et~al.(2012)Ohlmann, LaCasce, Washburn,
    Mariano, and
  Emery}]{Ohlmann2012}
Ohlmann, J.~C., LaCasce, J.~H., Washburn, L., Mariano, A.~J.,
and Emery, B.:
  Relative dispersion observations and trajectory modeling in the {S}anta
  {B}arbara {C}hannel, J. Geophys. Res., 117, C05\,040, 2012.

\bibitem[{Orfila et~al.(2005)Orfila, Jordi, Basterretxea,
    Vizoso, Marba,
  Duarte, Werner, and Tintor\'{e}}]{Orfila2005}
Orfila, A., Jordi, A., Basterretxea, G., Vizoso, G., Marba, N.,
Duarte, C.,
  Werner, F., and Tintor\'{e}, J.: Residence time and Posidonia oceanica in
  {C}abrera {A}rchipelago {N}ational {P}ark, Spain, Continental Shelf Research,
  25, 1339--1352, 2005.

\bibitem[{Pattanty\'{u}s-\'{A}brah\'{a}m
  et~al.(2008)Pattanty\'{u}s-\'{A}brah\'{a}m, T\'{e}l, Kr\'{a}mer, and
  J\'{o}zsa}]{Pattantyus2008}
Pattanty\'{u}s-\'{A}brah\'{a}m, M., T\'{e}l, T., Kr\'{a}mer,
T., and J\'{o}zsa,
  J.: Mixing properties of a shallow basin due to wind-induced chaotic flow,
  Advances in Water Resources, 31, 525--534, 2008.

\bibitem[{Ramis and Alonso(1988)}]{Ramis1988} Ramis, C. and
    Alonso, S.: {S}ea {B}reeze convergence line in {M}allorca.
    A
  satellite observation, Weather, 43, 288--293, 1988.

\bibitem[{Ramis and Romero(1995)}]{Ramis1995} Ramis, C. and
    Romero, R.: A {F}irst {N}umerical {S}imulation of the
  {D}evelopment and {S}tructure of the {S}ea {B}reeze in the {I}sland of
  {M}allorca, Ann. Geophy., 13, 981--994, 1995.

\bibitem[{Rypina et~al.(2011)Rypina, Scott, Pratt, and Brown}]{Rypina2011}
Rypina, I., Scott, S.~E., Pratt, L.~J., and Brown, M.~G.: Investigating the
  connection between complexity of isolated trajectories and {L}agrangian
  coherent structures, Nonlin. Processes Geophys, 18, 977--987, 2011.
  
\bibitem[{Shadden et~al.(2005)Shadden, Lekien, and
    Marsden}]{Shadden2005} Shadden, S.~C., Lekien, F., and
    Marsden, J.~E.: Definition and properties of
  Lagrangian coherent structures from finite-time Lyapunov exponents in two
  dimensional aperiodic flows, Physica D, 212, 271--304, 2005.

\bibitem[{Shadden et~al.(2009)Shadden, Lekien, Paduan, Chavez,
    and
  Marsden}]{Shadden2009}
Shadden, S.~C., Lekien, F., Paduan, J.~D., Chavez, F.~P., and
Marsden, J.~E.:
  The correlation between surface drifters and coherent structures based on
  high-frequency radar data in Monterey Bay, Deep Sea Research Part II: Topical
  Studies in Oceanography, 56, 161 -- 172, 2009.

\bibitem[{Song and Haidvogel(1994)}]{Song1994} Song, Y. and
    Haidvogel, D.: A {S}emi-implict {O}cean {C}irculation
    {M}odel
  {U}sing a {G}eneralized {T}opography-{F}ollowing {C}oordinate {S}ystem,
  Journal of Computational Physics, 115(1), 228--244, 1994.

\bibitem[{T\'el and Gruiz(2006)}]{TelGruiz2006} T\'el, T. and
    Gruiz, M.: Chaotic dynamics: {A}n introduction based on
    classical
  mechanics, Cambridge Univ. Press, Cambridge, 2006.

\bibitem[{{Tew Kai} et~al.(2009){Tew Kai}, Rossi, Sudre,
    Weimerskirch, L\'opez,
  Hern\'andez-Garc\'ia, Marsac, and Gar\c{c}on}]{TewKai2009}
{Tew Kai}, E., Rossi, V., Sudre, J., Weimerskirch, H., L\'opez,
C.,
  Hern\'andez-Garc\'ia, E., Marsac, F., and Gar\c{c}on, V.: Top marine
  predators track {L}agrangian coherent structures, Proceedings of the National
  Academy of Sciencies of the USA, 106, 8245--8250, 2009.

\bibitem[{Tudur{\'\i} and Ramis(1997)}]{Tuduri1997}
    Tudur{\'\i}, E. and Ramis, C.: The environments of
    significant convective
  events in the western Mediterranean, Weather and forecasting, 12, 294--306,
  1997.

\end{thebibliography}
\end{document}